\documentclass[twocolumn, preprint]{aastex631}
\usepackage{subfigure}
\usepackage{color}
\usepackage{xcolor}
\usepackage{amsmath}
\usepackage{mathrsfs}
\usepackage{multirow}
\usepackage{soul}
\usepackage{amssymb}
\usepackage{hyperref}
\usepackage{bm}

\newcommand{\unit}[1]{\mathrm{\ #1}}

\def\be{\begin{equation}}
\def\ee{\end{equation}}
\def\bd{\begin{displaymath}}
\def\ed{\end{displaymath}}
\def\ba{\begin{aligned}}
\def\ea{\end{aligned}}
\def\bh{M_{\bullet}}

\def\msun{M_{\odot}}

\begin{document}

 \title{The Secular Periodic Evolution of X-ray Quasi-periodic Eruptions Driven by Star-disc Collisions}
\author{Jiajun Xian}
\affiliation{School of Physics and Materials Science, 
Guangzhou University, 
510006 Guangzhou, China}
\author{Fupeng Zhang}
\correspondingauthor{Fupeng Zhang}
\email{zhangfupeng@gzhu.edu.cn}
\affiliation{School of Physics and Materials Science, 
Guangzhou University, 
510006 Guangzhou, China}
\affiliation{Key Laboratory for Astronomical Observation and Technology of Guangzhou, 510006 Guangzhou, China}
\affiliation{Astronomy Science and Technology Research Laboratory of Department of Education of Guangdong Province, Guangzhou 510006, China}
\author{Liming Dou}
\affiliation{School of Physics and Materials Science, 
Guangzhou University, 
510006 Guangzhou, China}
\affiliation{Key Laboratory for Astronomical Observation and Technology of Guangzhou, 510006 Guangzhou, China}
\affiliation{Astronomy Science and Technology Research Laboratory of Department of Education of Guangdong Province, Guangzhou 510006, China}
\author{Zhining Chen}
\affiliation{School of Physics and Materials Science, 
Guangzhou University, 
510006 Guangzhou, China}

 \begin{abstract}
 We study the secular periodic evolution of quasi-periodic eruptions (QPEs) for GSN069 and eRO-QPE2 assuming that they are driven by 
 star-disc collisions. We set up numerical simulations and compared them with the observed periodic decay of 
 $\sim -3160\pm720$\,s\,yr$^{-1}$ in 
 GSN069 and $\sim -370\pm40$\,s\,yr$^{-1}$ in eRO-QPE2. We find that: (1) Stellar mass black holes are unlikely  the orbiters in these two sources, as their periodic decay 
 are on the order of $<10$ s\,yr$^{-1}$; (2) A naked degenerate core (including white dwarf) is unlikely the orbiter in GSN069, as the decay is on the order of $<200$\,s\,yr$^{-1}$. However, it is possible in eRO-QPE2, although the required surface density of the accretion disc is relatively high (e.g., $\Sigma\gtrsim10^7\sim 10^8$\,g\,cm$^{-2}$); (3) Both the orbiters in GSN069 and eRO-QPE2 can be solar-like main-sequence stars (MSs). However, each collision can lead to gradual ablation of the stellar envelope in the order of $10^{-5}\sim 10^{-3}\msun$. To reproduce the observed decay while surviving for $\gtrsim 3$\,yr,
 the surface density of the disc needs to be within a certain range. For example, given a $1\msun$ MS orbiter
 the surface density of the disc gas should be in the range of $3\times10^5\sim 2\times10^6$g\,cm$^{-2}$ for GSN069 or $5\times10^4\sim 10^6$\,g\,cm$^{-2}$ for eRO-QPE2. 
 In both of these two sources, the MS can not {survive} for more than $\sim 12$\, yr. We expect that future observations of these two sources can help to distinguish whether the orbiters are degenerated compact objects or gaseous stars.  
 \end{abstract}

\section{Introduction}
X-ray quasi-periodic eruptions (QPEs) are sources that show intensive and repeating bursts of soft X-ray flares in galactic nuclei. 
So far several sources are already identified as QPEs, e.g., {
GSN 069~\citep{2019Natur.573..381M}, J1301.9+2747~\citep{2020A&A...636L...2G}, 
eRO-QPE1, eRO-QPE2~\citep{2021Natur.592..704A}, eRO-QPE3,
eRO-QPE4~\citep{Arcodia2024A&A...684A..64A}, and AT2019qiz~\citep{Nicholl2024},} 
all of which show regular/irregular patterns of recurring flares every several hours or days, 
and are persisting over time spans of several months to years. 
{Some of the X-ray sources can be possible candidates of QPE, 
e.g., Swift J023017.0+283603~\citep{Guolo2024NatAs...8..347G,Evans2023NatAs...7.1368E}, 
XMMSL1 J024916.6-041244~\citep{2021ApJ...921L..40C} and AT 2019vcb\citep{Quintin2023A&A...675A.152Q}}. 

There are increasing evidences that QPE sources may have been tightly linked to tidal disruption events (TDEs). For example, 
both GSN069~\citep{2019Natur.573..381M,2018ApJ...857L..16S} and {eRO-QPE3~\citep{Arcodia2024A&A...684A..64A}}
show decaying quiescence flux along with QPEs, a long-term characteristic of TDE.
The optical spectral analysis also suggests that GSN069 could have experienced a previous partial
TDE~\citep{2021ApJ...920L..25S,2022A&A...661A..55Z}. Before the onset of QPE, AT2019qiz, Swift J023017.0 + 283603, and AT 2019vcb were identified as TDEs~\citep{Nicholl2024,2021ApJ...921L..40C,Quintin2023A&A...675A.152Q}. 
All confirmed or candidate QPE sources host small massive black holes (MBH) with masses
ranging from $10^5\sim10^7\msun$\citep{Arcodia2024A&A...684A..64A,1983Ap&SS..95...11Z,2022A&A...659L...2W,Arcodia2024A&A...684A..64A,
Guolo2024NatAs...8..347G,Quintin2023A&A...675A.152Q}, which are consistent with the expected mass of MBH ($\lesssim10^8\msun$) in TDE scenarios. 

So far, there are ongoing debates of the underlying physical mechanism that drives the QPEs. The proposed models include
instabilities of the accretion disk~\citep{2021ApJ...909...82R,2022ApJ...928L..18P,2023ApJ...952...32P,2023MNRAS.524.1269K}; 
lensing of the accretion disks of a massive black hole binary~\citep{2021MNRAS.503.1703I}; and
mass transfer from an orbiting object onto the central MBH each time it passes the orbital pericenter~\citep{2020MNRAS.493L.120K,2022MNRAS.515.4344K,
2022ApJ...933..225W,2022A&A...661A..55Z,2024A&A...682L..14W,2022ApJ...941...24K,2023MNRAS.520L..63K,2023ApJ...947...32C,
2023ApJ...945...86L,2023MNRAS.524.6247L,2022ApJ...926..101M,2024arXiv241005166V}.
The model of collisions between an orbiter and the accretion disc surrounding an MBH (hereafter we called it ``star-disc collision" 
model~\footnote{The orbiter could be a main-sequence star, a compact object such as a stellar-mass black hole, a white dwarf, 
or a stellar core after a partial tidal disruption event. Nevertheless, we call the model as ``star-disc collision'' in the paper.}) has drawn much attention recently~\citep{2021ApJ...917...43S,2021ApJ...921L..32X,2023A&A...675A.100F,
2023ApJ...957...34L,2023MNRAS.526...69T,2024ApJ...973..101L,2024PhRvD.109j3031Z,
2024PhRvD.110h3019Z,2024arXiv240208722C,2024arXiv241005166V}. 

In the star-disc collision model, QPEs can be explained by the emission of heated gas {debris} 
resulting from violent impacts of the orbiter on the
accretion disc~\citep{1983Ap&SS..95...11Z,1998ApJ...507..131I,1999PASJ...51..571S,2004A&A...413..173N,2016MNRAS.457.1145P,2021ApJ...917...43S,
2023MNRAS.526...69T,2024arXiv241005166V,2024arXiv240714578Y}. Injections of the energies from the orbiter to the
disc gases may further evolve and affect the stability of the disc structure~\citep{2024ApJ...973..101L}. 

As the orbiter crosses the disc twice per orbit, it can produce alternative long and short recurring timing patterns due to the small eccentricity, the Schwarzschild or spin precession of the relativistic orbit~\citep[e.g.,][]{2021ApJ...921L..32X} 
and the precession of the disc plane if it is misaligned with the equatorial plane of MBH~\citep[e.g.,][]{2023A&A...675A.100F,2024PhRvD.109j3031Z,2024PhRvD.110h3019Z}.
Such timing patterns have already been found to be consistent with the observations of some QPE sources and some parameters of the system can be extracted~\citep{2021ApJ...921L..32X,2024PhRvD.109j3031Z,2024PhRvD.110h3019Z,2023A&A...675A.100F}.

There is an interesting possibility that the orbiter can be an extreme mass ratio inspiral (EMRI)~\citep{2021Natur.592..704A},
and thus can potentially be important candidates for gravitational wave sources for space-born gravitational wave
telescopes~\citep[e.g.,][]{2022ApJ...930..122C,Kejriwal2024}.
It is also proposed that the spin of MBH can be constrained by accurate observations of the timing of flares~\citep{2021ApJ...921L..32X}.

It is estimated that the lifetime of QPEs could range from several years to $10^{3}$ yr in a star-disc collision model, mainly
due to its gradual dissipation of momentum by colliding with the disc gas~\citep{1995MNRAS.275..628R,2021ApJ...921L..32X} or
the ablations of the envelope if the orbiter is a main-sequence (MS) or an evolved star~\citep{2020ApJ...889...94M,2023ApJ...957...34L,2024arXiv240714578Y}.
Thus, it is reasonable to expect that there is some secular evolution of the observed timing (periods, in particular) 
of these QPEs sources, of which the evolution may depend on the type of the orbiters (compact objects or MS). 
Among all QPE sources currently detected,
GSN069 and eRO-QPE2 show evidences of decaying orbital periods of QPEs over an observational span of $\sim 3.5$ years~\citep{2023A&A...670A..93M,Pasham2024}. 
In this study, we investigate whether the long-term periodic evolution of the QPE flares in GSN069 and eRO-QPE2 can be reproduced given different types of orbiters.

Considering the interesting connections between QPE and TDEs, 
we assume that the accretion disc is formed by the gas stripped from the (partial) TDE. 
The method adopted is based on our previous work~\citep{2021ApJ...921L..32X}, where
we adopt a full relativistic numeric method~\citep[developed in][]{2015ApJ...809..127Z,2017ApJ...849...33Z} to simulate the orbit and
obtain the timing of the QPE events. In addition, we include the drag forces from the gas on the orbiter
each time it crosses the disc plane. For stars, we also additionally consider the ablation of its envelope
in each collision. By comparing the simulated periodic decay with those of the observations, we can discuss the possibilities of the orbiter being stellar mass black holes (SBHs), compact degenerate objects, or MSs.

The paper is organized as follows. In Section~\ref{sec:summary_ob} we first summarize some of the important
properties of GSN069 and eRO-QPE2, including the long-term evolution of their QPEs. Section~\ref{sec:model_method} 
describes the details of the accretion disc models adopted, how to simulate the timing of QPEs, and the
evolution of the orbiter when the drag force of gas on accretion disc is considered. 
In Section~\ref{sec:result} we describe the simulation results under different types of orbiters and
discuss their consistencies with the observations of GSN069 and eRO-QPE2. 
Discussions of the results and conclusions are shown in Section~\ref{sec:discussion} and Section~\ref{sec:conclusion}, respectively.

\section{Summary of the QPEs in GSN069 and eRO-QPE2}
\label{sec:summary_ob}
\begin{table*}
    \caption{Summary of QPEs in GSN069}
    \centering
    \begin{tabular}{|c|c|c|c|c|}
    \hline
        Observation$^a$ & Date (start) & Time (day)$^b$ & $t_{rec}$ (s) & $P$ (s)$^c$ \\ \hline
        XMM3  & 2018-12-24 & $0 $& 29500 & $-$\\ 
        (0823680101)  & ~ & $0.34143$ & $-$ & $-$\\ 
        XMM4  & 2019-01-16 & $22.82463$ & 32850 & 64400 \\ 
        (0831790701) & ~ & $23.20484$ & 31550 & 63650\\ 
        ~ & ~ & $23.57000$ & 32100 & 64100\\ 
        ~ & ~ & $23.94153$ & 32000 & $-$\\ 
        ~ & ~ & $24.31190$ & $-$ & $-$\\ 
        Chandra$^d$  & 2019-02-14 & $52.32349 $& 32625 & 64950\\ 
       (22096)& ~ & $52.70109$ & 32325 & $-$\\ 
        ~ & ~ & $53.07522$ & $-$ & $-$\\ 
        XMM5  & 2019-05-31 & $158.52193$ & 29750 & 63600\\ 
        (0851180401)& ~ & $158.86626$ & 33850 & 64900\\ 
        ~ & ~ & $159.25804$ & 31050 & 64050\\ 
        ~ & ~ & $159.61742$ & 33000 & $-$\\ 
        ~ & ~ & $159.99936$ & $-$ & $-$\\ 
        XMM6   & 2020-01-10 & $382.16154$ & 27200 & 56000\\ 
        (0864330101) & ~ & $382.47636$ & 28800 & 64049\\ 
        ~ & ~ & $382.80969$ & 35249 & $-$\\ 
        ~ & ~ & $383.21767$ & $-$ & $-$\\ 
        XMM12  & 2022-07-07 & $1291.30528$ & (34400) & (54000)\\ 
        (0913990201) & ~ & $1291.70343$ & 19600 & \\
         & ~ & $1291.93028$ & $-$ & \\ \hline
    \end{tabular}
    \tablecomments{
        $^a$ The pointed XMM-Newton or Chandra observations used in this work. The number in bracket is the observation ID. \\
        $^b$ Time of flare with respect to the first one. \\
         $^c$ The period of QPE, defined as the time interval between the center of this flare and that of the second successive flare.
          In XMM12 the period in bracket is speculated according to the two QPEs and the correlation between 
          the oscillation in quiescent flux and QPE~\citep{2023A&A...674L...1M}.\\
        $^d$ {The Chandra dataset is obtained by the Chandra X-ray Observatory contained 
        in~\dataset[DOI: 10.25574/cdc.376]{https://doi.org/10.25574/cdc.376}}
    }
    \label{tab:obs_data_gsn069}
\end{table*}

\begin{table*}
    \caption{Summary of QPEs in eRO-QPE2}
    \centering
    \begin{tabular}{|c|c|c|c|c|}
    \hline
        Observation & {Date (start)} & Time (day)  & $t_{rec}$ (s) & $P$ (s) \\ \hline
        XMM1  & 2020-08-06 & $0$  & 8949 & 17549\\ 
        (0872390101)& ~ & $0.10358$  & 8600& 17550\\ 
        ~ & ~ & $0.20312$  & 8950 & 17400\\ 
        ~ & ~ & $0.30671$  & 8450 & 17550\\ 
        ~ & ~ &$ 0.40451$  & 9100 & 17551\\
        ~ & ~ & $0.50983$  & 8451 & 17400\\
        ~ & ~ & $0.60764 $ & 8949 & 17300\\
        ~ & ~ & $0.71122 $ & 8351 & $-$\\
        ~ & ~ & $0.80787$  & $-$ & $-$\\
        XMM2 & 2022-02-06 & $548.91919$  & 8520 & 16880\\
        (0893810501) & ~ & $549.01780$  & 8360 & $-$\\
        ~ & ~ & $549.11456$  & $-$ & $-$\\
        XMM3  & 2022-06-21  & 684.04397 & 8160 & 16620\\
        (0883770201) & ~ & 684.13842 & 8460 & 16530\\
        ~ & ~ & 684.23634 & 8070 & 16260\\
        ~ & ~ & 684.32974 & 8190 & 16200\\
        ~ & ~ & 684.42453 & 8010 & $-$\\
        ~ & ~ & 684.51724 & $-$ & $-$\\
        XMM4  & 2023-12-08 & $1218.99634$   & 8100 & 16450\\
        (0931791301)& ~ & $1219.09009 $ & 8350 & 16450\\
        ~ & ~ & $1219.18673$  & 8100 & 16250\\
        ~ & ~ & $1219.28048$  & 8150 & 16200\\
        ~ & ~ & $1219.37481$  & 8050 & $-$\\
        ~ & ~ & $1219.46798$  & $-$ & $-$\\ \hline
    \end{tabular}
    \tablecomments{Similar to Table~\ref{tab:obs_data_gsn069} but for eRO-QPE2.
    }
    \label{tab:obs_data_eroqpe2}
\end{table*}

In the star-disc collision model, the period is defined as the time interval between successive collisions of the orbiter with the disc in the same direction, i.e., either upward or downward. As the star collides with the disc two times per period, observationally
we can define the period ($P$) of a given QPE as the time interval between its flare peak and that of the second successive QPE.

In this work, we focus on GSN069 and eRO-QPE2.
{Details of the data reduction and extraction of the XMM-Newton and Chandra lightcurves of these two sources 
can be found in Appendix~\ref{sec:data_process}}. 
We obtained the observed timing of QPEs and the orbital period derived from the lightcurves of these two sources and summarized the results in Tables~\ref{tab:obs_data_gsn069} and~\ref{tab:obs_data_eroqpe2}. 

Note that for GSN 069,  the period of $\sim 54\pm4$ ks of QPEs on July 2022 is speculated from
the two new QPEs and the correlation between QPE and the quasi-periodic oscillation in the quiescence flux~\citep{2023A&A...674L...1M}. Thus, conservatively, the period should be considered as a lower limit of the period at that moment. Nevertheless, in this study we mainly discuss the decay of periods
assuming that the measured value does not deviate significantly from the true period of QPEs at that moment.
We also add discussions in Section~\ref{sec:discussion} of how such an assumption affects our conclusions.

The decay of the orbital period can be described by its relative change ($dP$) within some
observational time ($dt$), i.e. $\dot P=d P/dt$, which is a dimensionless quantity. 
From Table~\ref{tab:obs_data_gsn069}, by using the least square method
we obtain that for GSN069 over $\sim 3.5$ yr the decay rate is roughly 
{$\dot P\sim -3160\pm 720$\,s per year}.
For eRO-QPE2 (Table~\ref{tab:obs_data_eroqpe2}), 
the rate is approximately $\dot{P}\sim-370\pm40 $\,s per year. 

The bolometric luminosity of the disc in GSN069 varies mainly between
$0.2-6\times 10^{43}$ erg s$^{-1}$~\citep{2023A&A...670A..93M}.
The MBH mass estimation in GSN069 ranges from
$2\times10^5-10^6\msun$~\citep{2021ApJ...921L..32X,2019Natur.573..381M,
2023A&A...670A..93M,2024PhRvD.109j3031Z,2022A&A...659L...2W}. If we assume that the maximum luminosity of GSN069 is its Eddington luminosity, then the indicated mass of MBH should be $\sim 5\times10^5\msun$. Given $\bh=3\times10^5\msun$, 
the observed luminosity results suggest that the Eddington ratio of accretion varies between $0.1\sim1.7$. 

If adopting $\bh\sim 3\times10^5\msun$ in GSN069, 
the orbital semimajor axis and eccentricity of the orbiter are $310\sim 365r_g$ and $e=0.05\sim 0.1$, 
respectively~\citep{2021ApJ...921L..32X,2024PhRvD.110h3019Z} , 
where $r_g=G\bh /c^2$ is the gravitational radius of the MBH, $G$ and $c$ is the 
gravitational constant and velocity of the light, respectively.
The semimajor axis becomes $\sim 160r_g$ if adopting 
$\bh=10^6\msun$~\citep{2023A&A...675A.100F,2024PhRvD.109j3031Z}.

eRO-QPE2 is observed in the QPE-active phase from Aug.$2020$ to Dec.$2023$. 
During this phase, eRO-QPE2 preserves almost constant flux of the
disk component~\citep{2024A&A...690A..80A}. The disk luminosity is about 
$\sim10^{41}$erg $s^{-1}$ in \citet{2021Natur.592..704A} but is updated 
to $1.3\sim 4.3\times10^{43}$erg s$^{-1}$ by including a larger column 
densities~\citep{2024A&A...690A..80A}. The central mass 
of MBH in eRO-QPE2 is estimated to be $10^5\msun$ using stellar 
velocity dispersions~\citep{2022A&A...659L...2W}. 
These results suggest an accretion Eddington ratio of 
$1\sim3$ for eRO-QPE2. In the scenario of star-disc collision for QPEs, 
the orbital semimajor axis and eccentricity of the orbiter 
in eRO-QPE2 is estimated to be $ 300\sim 320r_g$ 
and $e=0.01\sim 0.06$~\citep{2024PhRvD.110h3019Z,2023A&A...675A.100F}.

\section{Numerical Method}
\label{sec:model_method}

\subsection{The Accretion Disk Model}
\label{sec:disk model}
\begin{figure}
\centering
    \includegraphics[scale=0.7]{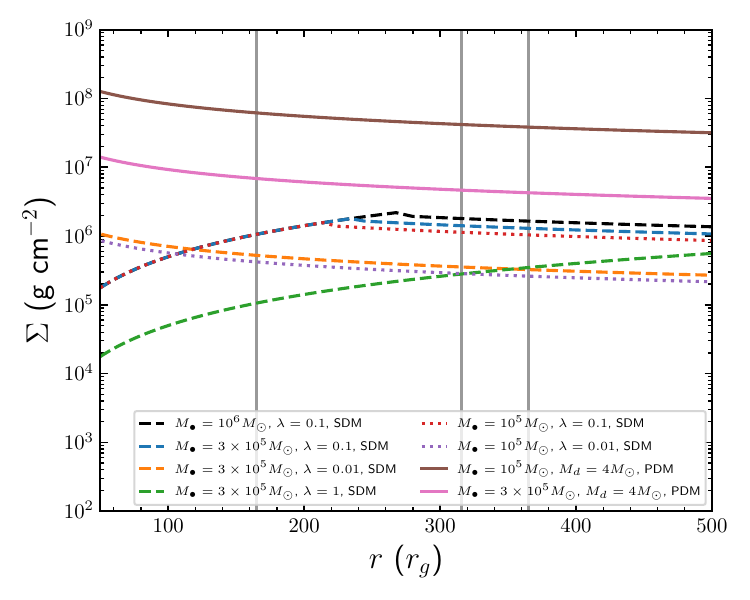}
    \caption{The surface density as a function of radius in 
    different accretion disc models. Solid lines represent the 
    profiles from the power law disk model (PDM) and dashed or dotted lines represent the 
    profiles from the standard disc model (SDM). From left to right, the three vertical light gray lines show the 
    reference position of $164.5\,r_g$ (orbiter for GSN069 assuming $\bh=10^6\msun$), 
    $316\,r_g$ (orbiter for eRO-QPE2 assuming $\bh=10^5\msun$) 
    and $365\,r_g$ (orbiter for GSN069 assuming $\bh=3\times10^5\msun$), respectively.}
    \label{fig:par_disk}
\end{figure}
To cover the large uncertainties of accretion physics and the environments 
around the central MBH, 
here we adopt two different accretion disk models. The first is the standard thin 
disk model (hereafter SDM) by~\citet{1973A&A....24..337S} (see
also \citet{2013LRR....16....1A} for a review),
which has been widely accepted for accretion of gaseous materials
around MBH. 
As typically
the orbiters of GSN069 and eRO-QPE2 intersect the disc at a distance of
$r=160-370r_{\rm g}$, we simply ignore any relativistic terms in SDM. Then the orbiters collide with the gas of the disc 
in the inner or middle regions in SDM. In the inner regions, 
the surface density $\Sigma$ can be given by~\citep{2013LRR....16....1A},
\begin{eqnarray}
\Sigma &=& [5 \unit{g~cm^{-2}}](\alpha^{-1} {\dot
m}^{-1})\,r^{3/2}. 
\label{eq:sigma_SDM2}
\end{eqnarray}
where $\alpha$ is the viscosity prescription of \citet{1973A&A....24..337S}, 
 $r$ is the distance to the MBH in unit of $r_g$. 
${\dot m}$ is the dimensionless accretion rate given by ${\dot m} = {\dot M} c^2/L_{\mathrm{Edd}}$, 
where $\dot M$ is the mass accretion rate, $L_{\rm Edd}=\epsilon \dot M_{\rm Edd}c^2$ is the
Eddington luminosity, $\dot M_{\rm edd}$ is the Eddington mass accretion rate,
$\epsilon$ is the radiative efficiency. In the middle region, we have~\citep{2013LRR....16....1A}
\begin{eqnarray}
\Sigma &=& [9 \times 10^4 \unit{g~cm^{-2}}](\alpha^{-4/5}
m^{1/5} {\dot m}^{3/5})\,r^{-3/5},  
\label{eq:sigma_SDM3}
\end{eqnarray} 
where $m = M_{\bullet }/M_{\odot}$.
The boundary between the inner and middle region is at $r\sim 112.7 (\alpha m)^{2/21}\dot m^{16/21}$. 
Throughout the paper, we assume that $\alpha=0.01$ and $\epsilon=0.1$.
Eddington ratio can be defined as $\lambda=\dot M/\dot M_{\rm edd}=\dot m\epsilon$.


The second model assumes that the surface density of the disc is a simple power-law function of radial distance, 
i.e., 
\be
\Sigma = \Sigma_0r^{-p}
\label{eq:sigma_PDM}
\ee
where
\be
\Sigma_0 = \frac{M_d(2-p)}{2\pi r_{\rm g}^2}\left(r_{\rm out}^{2-p}-r_{\rm in}^{2-p}\right)^{-1} 
\ee
where $M_d$ is the mass of the disk, $r_{\rm out}$ and $r_{\rm in}=6r_g$ are the dimensionless radius at the outer and inner edge 
of the disk respectively, $p$ is the power law index. Here we refer to the model as the power law disc model (PDM).

PDM has been commonly adopted for the study of the disc in
TDEs~\citep{2014ApJ...784...87S,2016MNRAS.455.1946F,2023A&A...675A.100F,2012PhRvL.108f1302S}.
However, the exact value of $p$ is so far unclear. \citet{2012PhRvL.108f1302S} adopts $p=-3/2\sim 1$. 
\citet{2016MNRAS.455.1946F} adopts $p=3/5$ based on
an analytical slim disc model~\citep{1997astro.ph..8265U,2009MNRAS.400.2070S}. 
Suggested by numerical simulations that the surface density of the disc shortly after the tidal disruption is a slow decreasing function of the radius~\citep{2011ApJ...736..126M}, here we adopt $p=3/5$, following the same value of those in
\citet{2016MNRAS.455.1946F}. Although $p$ could vary during different phases of the 
tidal disruption events~\citep{2008MNRAS.390..781M,2014ApJ...784...87S}, here we simply assume that it does not vary during the
span of interest ($\sim 3$yr). In the TDE scenario, it is natural to expect that
the disc mass $M_d$ is about a fraction ($\lesssim 50\%$) of the mass of the disrupted star.

In PDM, the surface density of the disc is regulated by the mass of the disk. 
As the disc in PDM is expected to be fed by the stream of gas of the disrupted
star, its density can be quite high, and the corresponding Eddington accretion rate
$\lambda$ can be much higher than unity. 

The outer edge of the disk can be about $2\sim 6$ times of the tidal radius of the
star~\citep{2016MNRAS.455.1946F,2023A&A...675A.100F}, which is given by
\be
R_{t}=\left(\frac{M_{\bullet } }{M_{\star  }} \right)^{1/3} R_{\star }.
\ee 
For GSN069 and eRO-QPE2, for solar mass stars $R_t=47\sim 200\,r_{\rm g}$.
As the orbiter of the two sources is located at a distance of $160\sim 370\,r_g$, here we fix
$r_{\rm out}=500 r_g$. A slightly different value of $r_{\rm out}$ affects only the normalization rather than the density profile, and thus can be absorbed by using a different value of the mass of the accretion disc.

The typical disc profiles as a function of the distance from the MBH in these two models are shown in Figure~\ref{fig:par_disk}. For orbiters (located in $160-370r_g$) in GSN069 and eRO-QPE2, 
the surface density is generally in the order of $10^5\sim10^8$\,g\,cm$^{-2}$. 
The typical slope index of the density is of $-3/5$ for
the case of SDM with $\lambda<1$. However, the slope index changes to $3/2$ if $\lambda=1$.


\subsection{Numerical simulations of the orbiter}
\label{subsec:num_orbiter}
Similarly to~\citet{2021ApJ...921L..32X}, we adopt the full relativistic numerical method previously developed in~\citet{2015ApJ...809..127Z} (see also \citet{2017ApJ...849...33Z}) to obtain the geodesics of the orbiter
around the MBH (when it does not collide with the disk). 
{In general,  for Kerr metric the motion of non-zero-mass particles (such as a star or compact object) or a photon can be controlled by the following equations 
in the Boyer-Lindquist coordinate $(r,\theta,\phi,t)$ (Adopting units $G=c=\bh=1$):
}

\be
\ba
\Sigma \dot r&=\pm \sqrt{R}\\
\Sigma \dot \theta&=\pm \sqrt{\Theta}\\
\Sigma \dot \phi&=-\chi+\zeta/\sin^2\theta+\chi T/\Delta\\
\Sigma \dot t&=-\chi^2\sin^2\theta+\chi\zeta+(r^2+\chi^2)T/\Delta\\
\ea
\ee
where 
\be
\ba
R&=(1-\xi^2)r^4+2\xi^2r^3+[\chi^2(1-\xi^2)-q^2-\zeta^2]r^2,\\
&+2[(\chi-\zeta)^2+q^2]r-\chi^2q^2,\\
\Theta &= q^2-[\chi^2(\xi^2-1)+\zeta^2/\sin^2\theta]\cos^2\theta,\\
\Delta&=r^2-2r+\chi^2,\\
\Sigma&=r^2+\chi^2\cos^2\theta,\\
T&=r^2+\chi^2-\zeta \chi,
\ea
\ee
{$\xi=m/E$, $\zeta=L_z/E$, and $q^2=Q/E^2$. 
$m$ is the mass of the orbiter ($m=0$ for photons). The energy at infinity $E$,
azimuthal angular momentum  $L_z$ and Carter's constant $Q$ are the constants of motion that is determined by the 
initial condition of the orbiter. $\chi$ is the dimensionless spin parameter of MBH.}

{We adopt the explict fifth(fourth)-order Runge Kutta method DORPI5~\citep{DORMAND198019} to integrate the equations of 
motion of a star or a compact object. }
The propagation of photons from the QPE flaring event to a distant observer is accurately calculated
by Jacobian elliptic functions and the Gauss-Kronrod integration scheme.
{More calculatioin details of the trajectories of 
a star/compact object and a photon 
see Section 3 and 4 of \citet{2015ApJ...809..127Z}, respectively.}

{
Similar to~\citet{2015ApJ...809..127Z}, the initial values of the constants of motion, i.e., 
$\xi$, $\zeta$ and $q$ is derived by the three velocity of the orbiter, i.e., 
$\vec v_{\star} (v^i_{\star}$, for $i=r, \theta,\phi$) obtained 
in the spherical coordinates of the local non-rotating rest frame (LNRF). 
The initial three velocity is determined according to the initial instantaneous
 Keplerian orbital elements of the orbiter: semimajor axis ($a$), eccentricity ($e$),
orbital inclination ($I$), ascending node ($\Omega$), argument of periapsis ($\omega$) 
and true anomaly ($f$). On the controversy, at any given time of integration, we can obtain the 
three velocity of the orbiter from its relativistic four velocity ($u_\star^\alpha$, $\alpha=t,r,\theta,\phi$) in 
Boyer-Lindquist coordinates~\citep{1972ApJ...178..347B,1973blho.conf..215B,1973grav.book.....M}
using Equation 9-11 of~\citet{2015ApJ...809..127Z}. }

In general, we can obtain the orbits of QPEs given some spinning of the MBH, as our numerical method adopts the Kerr metric. 
However, we find that including the effect of spin leads to only the difference in the oscillation patterns of the timing of QPE and does not affect the long-term orbital decay of the periods. Thus, for simplicity, here we assume that the MBH is non-spinning (a Schwarzschild MBH). We do not consider the tilt of the accretion disc relative to the equatorial plane of the MBH, although we notice that
the Lense-Thirring orbital precession can also lead to some additional
oscillation patterns of the timing of QPE~\citep{2023A&A...675A.100F}.

We assume that the QPE flares are generated when the orbiter crosses the middle plane of the accretion disk.
Given the coordinate time of the star at the crossing $t_{\star }$, and the propagation time of the flare to a distance observer $t_{\rm prop}$, we
obtain the time of arrival (TOA) $t_{\rm TOA} $ of each QPE by $t_{\rm TOA} =t_{\star } +t_{\rm prop}$.
We consider only the photons of flares propagating in the shortest distance to the observer (the primary image), 
although in principle there can be more than one image of the flare due to the gravitational lensing (the photons bend over the back of the MBH before reaching to the observer). However, the intensity of these high-order images of flares is usually much lower than
the primary one if the flares occur at distance $r\gg 10r_g$~\citep{1994ApJ...422..208K,2010MNRAS.402.1614D}. 

\subsection{The drag force felt by the orbiter}
\label{sec:change of velocity and mass strip}
The orbiter feels a drag force ($F_{\rm drag}$) each time it crosses the disc plane, due to the impact between the gas on the disc and the orbiter. 
One of the possible sources of the drag force is the gravitational drag of the gas~\citep{1980ApJ...240...20R}
\be
F_{\rm drag,grav} \approx \frac{4\pi\ln \Lambda(GM_{\star })^{2}\rho}{\left | v_{\rm rel}\right |^{2}} 
\label{eq:fdrag_grav}
\ee
where $\Lambda=b_{\rm max}/b_{\rm min}$ is the Coulomb factor, $b_{\rm max}$ ($b_{\rm min}$) is the maximum (minimum) impact parameter associated with the interaction. Approximately $b_{\rm max}\sim2H/\sin I$ and $b_{\rm min}\sim R_\star$
where $H$ is the scale height, $I$ is the inclination angle of the orbiter and $R_\star$ is the
physical size of the orbiter. {$\vec v_{\rm rel}=\vec v_\star-\vec v_{\rm disk}$ is the
relative velocity between the orbiter $(\vec v_\star)$ and the gas on the disc ($\vec v_{\rm disk}$).}

Another source of drag force is the friction of the gas on the surface of a star
\be
F_{\rm drag,fri}\approx Q\pi R_{\star}^{2}\rho\left| v_{\rm rel} \right |^{2}, 
\label{eq:fdrag_geo}
\ee
where $Q$ is the drag efficiency. Here we simply fix $Q=1$. 
$\rho$ is the density of the gas,  $R_\star$ is the geometric size of the orbiter, which is the physical radius of a star or the event horizon of an SBH.

If the orbiter is an SBH, we have $v_{\rm rel}^2\ll G M_\star/R_\star$ and thus 
$F_{\rm drag,grav}$ dominate over $F_{\rm drag,fri}$. 
In this case, we find $\ln \Lambda\sim 10$ for both orbiters in GSN069 and eRO-QPE2. 
For simplicity, here we fix $\ln\Lambda=10$.
In other cases, e.g., MS stars or other degenerate compact stars, e.g., white dwarfs (WDs), 
we always have $F_{\rm drag,fri}\gg F_{\rm drag,grav}$. 

The change of the  velocity of the orbiter per collision is then given by
\be
\Delta \vec {v} =-\frac{F_{\rm drag}}{M_{\star}} 
\frac{2H}{v_{\rm \star,z}} \frac{\vec v_{\rm rel} }{\left | v_{\rm rel}\right | } 
\label{eq:change_v}
\ee
where $v_{\rm \star,z}$ is the vertical velocity of the orbiter and $H$ is the scale height of the disc.
Since the drag forces are proportional to the density $\rho$, $\Delta \vec{v}$ is proportional to
$\propto \rho H\propto \Sigma$ and thus independent with the scale height of the disc.

For simplicity, the gas on the disc is assumed to follow pure circular orbits on the equatorial plane of the MBH ($\theta=\pi/2$), 
and thus it has only the non-zero component in the $\phi$ direction. In LNRF and for a Schwarzschild black hole, it is given by~\citep{1973blho.conf..343N,1972ApJ...178..347B}:
\be
v_{\rm disk}=v_{\phi }= \sqrt{\frac{G\bh}{r}}\left(1-2\frac{G\bh}{rc^2}\right)^{-1/2}
\ee

According to $\vec v_\star$ and $\vec v_{\rm disk}$ above, we can calculate the change of velocity of the star after each collision. Then we transform the velocity 
back to its relativistic four velocity and calculate its geodesics around the MBH again according to
procedures described in Section~\ref{subsec:num_orbiter}. By repeating such a process, we can obtain a series of QPE signals that arrive on the observer's frame, of which the periods can be used to compare with those observed in GSN069 and eRO-QPE2.

The average value of $\dot P$ within one period can be analytically estimated for near-circular orbits.
If $I$ is the orbital inclination of the orbiter with respect to the disc plane, then we have
\be
\dot P\simeq -\frac{3\sqrt{2} F_{\rm drag} (2H)}{M_\star v_\star^2} \frac{(1-\cos I)^{1/2}}{\sin I}.
\label{eq:dotp}
\ee
In the case of an SBH orbiter, 
the gravitational drag dominates, and from Equation~\ref{eq:dotp} we obtain that the orbital decay is given by 
\be\ba
\dot{P}_{\rm grav} &\sim -6\sqrt{2}  \pi \ln \Lambda\frac{M_{\star } }{M_{\bullet}^{2}  } a^{2} \Sigma\frac{(1-\cos I)^{-1/2}}{\sin I}\\
&\sim -8.5\times 10^{-4}  {\rm s/yr} \times\frac{M_{\star }}{M_{\odot  }}\frac{\Sigma}{10^{5}\rm g\,\rm cm^{-2}}\\
&\times \left(\frac{a}{300r_{\rm g} } \right)^{2}\frac{(1-\cos I)^{-1/2}}{\sin I}.
\label{eq:dotp_grav}
\ea\ee
For a WD or MS, the dominant mechanism of orbital decay is the frictional force of the gas, according to Equation~\ref{eq:fdrag_geo} we have 
\be\ba
\dot P_{\rm fri}\simeq &-\frac{6\sqrt{2} Q\pi R_\star^2 \Sigma }{M_\star}\frac{(1-\cos I)^{3/2}}{\sin I}\\
&\simeq -205 \,\mathrm{s/yr} \times Q\left(\frac{R_{\ast }}{R_{\odot } } \right)^{2}  
\frac{\Sigma}{10^{5}\rm g\,\rm cm^{-2}   } \\
 &\times\frac{M_{\odot}}{1 M_{\star }}\frac{(1-\cos I)^{3/2}}{\sin I}  
 \label{eq:dotp_fri}
\ea\ee

\section{The simulated long-term periodic evolution of QPEs}
\label{sec:result}

\begin{table*}
    \caption{Summary of simulations for GSN069}
    \centering
    \begin{tabular}{|c|c|c|c|c|c|c|c|c|c|c|c|c|}
    \hline
    Model & Orbiter &  $M_\star^a$ & $R_\star^b$ & $\bh^c$  &  disc$^d$ & $\Sigma(a_i)^e$  & {$I^f_i$}  & $\dot P^g$ (s/yr) & $\tau^h$ (day) & Con.?$^i$\\ 
    \hline
    SBH1  &\multirow{2}{*}{SBH }& \multirow{2}{*}{$10$ } &\multirow{2}{*}{$0.0000425$}  &\multirow{2}{*}{$3\times10^5$}  
    & PDM, $4\msun$ & $4.3\times10^6$ & $160^\circ$ & $-1.1$ & \multirow{4}{*}{$\infty$} & No\\
    \cline{1-1}\cline{6-9}\cline{11-11}
    SBH2&  &  &  &  &   SDM, $\lambda=0.1$ &  $1.3\times10^6$ & $160^\circ$ & $-0.3$ & & No\\
    \cline{1-9}\cline{11-11}
    WD1  &\multirow{2}{*}{WD} &\multirow{2}{*}{$0.6$} & \multirow{2}{*}{$0.012$} &\multirow{2}{*}{$3\times10^5$} &PDM, $4\msun$ & $4.3\times10^6$  & $160^\circ$ & $-17$ & & No\\
    \cline{1-1}\cline{6-9}\cline{11-11}
    WD2    &  &  &    &  &  SDM, $\lambda=0.1$  &  $1.3\times10^6$ & $160^\circ$ & $-5.2$ & & No\\
    \hline
    MSS51 &\multirow{14}{*}{MS}& \multirow{11}{*}{ $1$} &\multirow{11}{*}{$1$}  
    &\multirow{11}{*}{$3\times10^5$}  &\multirow{3}{*}{PDM, $4\msun$} &  \multirow{3}{*}{$4.3\times10^6$} & $45^\circ$ & $-997$ & $1299$ & No\\
    \cline{1-1} \cline{8-10}\cline{11-11}
        MSS52&  &  &  &  &           &  & $60^\circ$ & $-2730$ & $1036$ & No\\
        \cline{1-1}  \cline{8-10}\cline{11-11}
        MSS53&  &  &  &  &           &  & $145^\circ$ &  $-28343$  & $387$ & No\\
        \cline{1-1}  \cline{6-10}\cline{11-11}
        MSS54&  &  &  &  &  PDM, $2\msun$        & $2.2\times10^6$ & $92^\circ$ & $-3113$ & $1332$& Yes\\
        \cline{1-1}  \cline{6-10}\cline{11-11}
         MSS55&  &  &  &  &  \multirow{3}{*}{SDM, $\lambda=0.1$}  & \multirow{3}{*}{$1.3\times10^6$} & $45^\circ$ & $-350$ & $4693$ & No\\
         \cline{1-1}  \cline{8-10}\cline{11-11}
        MSS56&  &  &  &  &  &  & $120^\circ$ & $-3086$ & $1660$ & Yes\\
        \cline{1-1}  \cline{8-10}\cline{11-11}
        MSS57&  &  &  &  &   &  & $145^\circ$ & $-8619$ & $1278$ & No\\
        \cline{1-1}  \cline{6-10}\cline{11-11}
        MSS58&  &  &  &  &  \multirow{2}{*}{SDM, $\lambda$ varying}  & \multirow{2}{*}{$3\sim 6 \times10^5$} & $145^\circ$ & $-1229$ & $3720$ & No\\
        \cline{1-1}   \cline{8-10}\cline{11-11}
        MSS59&  &  &  &  &  &  & $159^\circ$ & $-3137$ & $3558$ & Yes\\
        \cline{1-1}\cline{5-10}\cline{11-11}
        MSS61&  &  &  &  \multirow{3}{*}{$10^6$}  & PDM, $4\msun$      & $6.2\times10^5$ & $146^\circ$ & $-2987$ & $1953$ & Yes\\
        \cline{1-1}\cline{6-10}\cline{11-11}
        MSS62&  &  &  &                          & SDM, $\lambda=0.1$  & $1.1\times10^6$ & $130^\circ$ & $-2905$ & $1585$ & Yes\\
        \cline{1-1}\cline{6-10}\cline{11-11}
        MSS63&  &  &  &                          & SDM, $\lambda$ varying  & $3\sim6\times10^5$ & $160^\circ$ & $-3020$ & $2839$ & Yes\\
    \cline{1-1}\cline{3-10}\cline{11-11}
    MSL1 & & \multirow{3}{*}{ $0.5$} &\multirow{3}{*}{$0.453$}  
    &\multirow{3}{*}{$3\times10^5$}  &\multirow{2}{*}{SDM, $\lambda=0.1$} &  \multirow{2}{*}{$1.3\times10^6$} & $145^\circ$  & $-9267$ & $923$ & Yes\\
    \cline{1-1}\cline{8-10}\cline{11-11}
    MSL2&  &  &  &                          &                         &                 & $80^\circ$ & $-1307$ & $1097$& No\\
    \cline{1-1}\cline{6-10}\cline{11-11}
    MSL3&  &  &  &                          & SDM, $\lambda$ varying  & $3\sim6\times10^5$ & $160^\circ$ & $-2805$ & $2773$ & Yes\\
    \hline
    HeCore51  &\multirow{5}{*}{He core} &
    \multirow{5}{*}{$0.26$} & \multirow{5}{*}{$0.03$} &\multirow{3}{*}{$3\times10^5$} &   PDM, $4\msun$  
      & $4.3\times10^6$  & $160^\circ$ & $-244$ & \multirow{5}{*}{$\infty$} & No\\
    \cline{1-1}\cline{6-9}\cline{11-11}
    HeCore52   &  &  &  &  &   SDM, $\lambda=0.1$  & $1.3\times10^6$ & $160^\circ$ & $-74$ &  & No\\     
    \cline{1-1}\cline{6-9}\cline{11-11}
        HeCore53   &  & & &  & SDM, $\lambda$ varying & $3\sim 6 \times10^5$ & $160^\circ$ & $-22$ &  & No\\
        \cline{1-1}\cline{5-9}\cline{11-11}   
    HeCore61  & & &   &\multirow{2}{*}{$10^6$} &      PDM, $4\msun$ & $6.2\times10^5$ & $160^\circ$ & $-36$ & & No\\
    \cline{1-1}\cline{6-9}\cline{11-11}
    HeCore62   &  &  &    &  & SDM, $\lambda=0.1$  & $1.1\times10^6$ &  $160^\circ$ & $-60$ & & No\\       
        \hline        
    \end{tabular}
    \tablecomments{
        $^a$  The (initial) mass of the orbiter, in units of $\msun$;\\
        $^b$  The (initial) radius of the orbiter, in units of $R_\odot$;\\
        $^c$  The mass of MBH, in a unit of $\msun$;\\
        $^d$  The model of the disc adopted; The mass after ``PDM'' means the mass of the accretion disc 
        assumed in the model.
        In SDM, $\lambda$ is the Eddington ratio adopted. ``$\lambda$ varying`` means that we vary the 
        Eddington ratio according to~\citet{2023A&A...670A..93M};  \\
        $^e$  The surface density of the disc (in unit of g\,cm$^{-2}$) at the start of the simulation; For the SDM models 
        adopt a varying Eddington ratio, the number shows the corresponding surface density 
        within the first $1290$ days. \\
        $^f$  The initial orbital inclination of the orbiter with respect to the disc plane;\\
        $^g$  The decay of orbital period within the first 1290 days;\\
        $^h$  The lifetime of the orbiter. For an MS orbiter, it is limited by the gradual ablation of its envelope 
        after each collision with the gas on the disc.\\
        $^i$ Whether the model consists with the observations of GSN069, i.e., 
        if the orbital decay is in the range of $3160\pm720$\,s\,yr$^{-1}$ and survival time $>3.5$ yr.
        }
    \label{tab:model_gsn069}
\end{table*}

\begin{table*}
    \caption{Summary of simulations for eRO-QPE2}
    \centering
    \begin{tabular}{|c|c|c|c|c|c|c|c|c|c|c|c|c|}
        \hline
        Model & Orbiter &  $M_\star$ & $R_\star$ & $\bh$  &  disc & $\Sigma(a_i)$  & $I_i$ &  $\dot P$ (s/yr) & $\tau$ & Con.?\\ 
        \hline
       SBH1  &\multirow{2}{*}{SBH }& \multirow{2}{*}{$10$ } &\multirow{2}{*}{$4.25\times10^{-5}$}  &\multirow{2}{*}{$10^5$}  
       & PDM, $4\msun$ & $4.1\times10^7$ & $160^\circ$  & $-8.2$ & \multirow{5}{*}{$\infty$} & No\\
       \cline{1-1}\cline{6-9}\cline{11-11}
       SBH2&  &  &  &  & SDM, $\lambda=0.1$ &$1.1\times10^6$ & $160^\circ$ &  $-0.22$ & & No\\
        \cline{1-9}\cline{11-11}
        WD1  &\multirow{3}{*}{WD} &\multirow{3}{*}{$0.6$} & \multirow{3}{*}{$0.012$} &\multirow{3}{*}{$10^5$} & PDM, $4\msun$ & $4\times10^7$  & $160^\circ$  & $-166$ & & No\\
        \cline{1-1}\cline{6-9}\cline{11-11}
         WD2 &  &  &  &  & PDM, $9\msun$ & $9\times10^7$ & $160^\circ$ &$ -376$ & & Yes\\
         \cline{1-1}\cline{6-9}\cline{11-11}
         WD3 &  &  &  &  & SDM, $\lambda=0.1$ & $1.1\times10^6$  & $160^\circ$ &$ -4.5$ & & No\\
        \hline
        MSS1 &\multirow{8}{*}{MS} &\multirow{6}{*}{$1$ } &\multirow{6}{*}{$1$} 
        &\multirow{7}{*}{$10^5$ }& PDM, $0.5\msun$ & $5\times10^6$ & $30^\circ$ &   $-562$ & $447$ & No\\
        \cline{1-1}    \cline{6-10}\cline{11-11}
        MSS2&  &  &  &  &  PDM, $0.01\msun$ & $10^5$ & $135^\circ$ &   $-380$ & $4536$ & Yes\\
        \cline{1-1}\cline{6-10}\cline{11-11}
        {MSS3}&  &  &  &  &  {PDM, $M_{d}$ varying} & $10^5$ & $141^\circ$ &   $-314$ & $>5000$ & Yes\\
        \cline{1-1}\cline{6-10}\cline{11-11}
        MSS4&  &  &  &  & SDM, $\lambda=0.01$ &  $2.8\times10^5$ & $95^\circ$ &   $-339$ & $2431$ & Yes\\
        \cline{1-1}\cline{6-10}\cline{11-11}
        MSS5&  &  &  &  & SDM, $\lambda=0.1$&  $1.1\times10^6$  &     $47^\circ$ & $-349$ & $1263$ & Yes\\
        \cline{1-1}\cline{6-10}\cline{11-11}
        MSS6&  &  &  &  & SDM, $\lambda=1$&  $2.8\times10^5$  &     $102^\circ$ & $-372$ & $2628$ & Yes\\
        \cline{1-1}\cline{3-4}\cline{6-10}\cline{11-11}
        MSL1 &  &\multirow{2}{*}{$0.5$ } &\multirow{2}{*}{$0.453$} 
        & & PDM, $0.01\msun$ & $10^5$ & $141^\circ$ &   $-380$ & $3009$ & Yes\\
        \cline{1-1}\cline{6-10}\cline{11-11}
        MSL2&  &  &  &  & SDM, $\lambda=1$&  $2.8\times10^5$  &  $98^\circ$ & $-361$ & $1727$ & Yes\\
           \hline
        HeCore1 &\multirow{5}{*}{He core} &\multirow{5}{*}{$0.26$} &\multirow{5}{*}{$0.03$} 
         &\multirow{5}{*}{$10^5$}  & \multirow{3}{*}{PDM, $4\msun$} &  \multirow{3}{*}{$4\times10^7$}  & $45^\circ$ & $-67$ 
         & \multirow{5}{*}{$\infty$} & No\\
         \cline{1-1}     \cline{8-9}\cline{11-11}
         HeCore2&  &  &  &    &  &   &$ 160^\circ$ & $-2554$ & & No \\
         \cline{1-1}\cline{8-9}\cline{11-11}
         HeCore3&  &  &  &    &                   &   &$ 100^\circ$ & $-390$ & & Yes\\
         \cline{1-1}\cline{6-9}\cline{11-11}
         HeCore4&  &  &  &    &  PDM, $1\msun$ & $10^7$ & $150^\circ$ & $-388$ & & Yes\\
         \cline{1-1}\cline{6-9}\cline{11-11}
         HeCore5&  &  &  &    & SDM, $\lambda=0.1$ & $1.1\times10^6$   &$ 160^\circ$ & $-65$ && No\\
         \hline
        \end{tabular}
    \tablecomments{Similar to Table~\ref{tab:model_gsn069} but for eRO-QPE2. 
    The orbital decay is measured within the first $1220$ days. Model is considered consist with 
    the observation of eRO-QPE2 if $\dot P$ is in the range of $370\pm40$\,s\,yr$^{-1}$ and the orbiter survives for more than $3.5$yr. 
    {Note that Model MSS3 adopt $M_d=0.01\msun$ 
    within the first $684$\,days (corresponding to the time of the third observation in Table~\ref{tab:obs_data_eroqpe2}) 
    and $M_d=2\times10^{-3}\msun$ after that. Although $\dot P=-314$\,s\,yr$^{-1}$ measured within 
    time span of observation seems lower than the required decay of period, 
    this model actually provides a better consistency with the observation of QPEs before and after Jun. 21, 2022 
    (See Figure~\ref{fig:ero_QPE2_mass_strip}). }
    }
    \label{tab:model_eroqpe2}
\end{table*}

In this section, we set up simulations according to Section~\ref{sec:model_method} and
study the long-term periodic evolution of QPEs in GSN069 and eRO-QPE2. 

The SDM parameters are adopted according to the observed features described in
Section~\ref{sec:summary_ob}. For GSN069 we can adopt $\lambda=0.1$ throughout all active phases,
or we consider a varying Eddington ratio according to the
model in~\citet{2023A&A...670A..93M}. We assume two possible masses of 
GSN069, i.e., either $\bh=3\times10^5\msun$ or $\bh=10^6\msun$.
For eRO-QPE2 we consider the case $\lambda=0.01$, $0.1$ or $1$ in 
SDM. 

For PDM here we consider mainly the disc profiles that are not covered by SDM. 
For example, for GSN069 we consider only the case in which
the disc mass $\ge 2\msun$, as it is quite similar to SDM if we adopt $M_d=1\msun$.
For eRO-QPE2 we consider the disc mass of $0.01\msun$, $0.5\msun$, or $4-10\msun$. 
The first mass is suggested by~\citet{2023A&A...675A.100F}. The rest are set so that the surface density are not covered by SDM. In principle, we can set the disc mass $>10\msun$, however, it is unlikely as it requires the progenitor star larger than $\gtrsim20\msun$.

We focus on four of the possible types of the orbiter, i.e., 
an SBH, a WD, an MS, and a stripped core of a red giant (a naked degenerate Helium core).
We assume that before the onset of the accretion disc by a tidal disruption event, 
these orbiters already exist with the initial orbits suggested by observations of
QPEs in GSN069 and eRO-QPE2. For GSN069, the initial orbital semimajor axis
is $a=365r_g$ if assuming $\bh=3\times10^5\msun$ and $a=164.5r_g$ if assuming $\bh=10^6\msun$. 
Similarly, $a=316r_{\rm g}$ for eRO-QPE2 assuming $\bh=10^5\msun$.
{We assume an initial orbital eccentricity of $0.05$ for GSN069.
For eRO-QPE2 the initial eccentricity is assumed to be $0.025$.
Adopting a slightly different initial value of eccentricity affects 
mainly the patterns of the recurring time of QPEs rather than the decay 
of the orbital period.
} 

We primarily focused on the cases that are more consistent with the observations. Such models
should satisfy at least the following three conditions: (1) The
orbiter should survive for at least $3.5$\,yr ($\sim 1300$\,days); (2) The orbital decay of $\sim -3160\pm720$\,s\, yr$^{-1}$
for GSN069 or $\sim -370\pm40$\,s\,yr$^{-1}$ for eRO-QPE2 should be reproduced; (3) 
The orbital inclinations should be in the range of $20^\circ<I<160^\circ$. 
If $I$ is close to zero or $180^\circ$ we are not expected to see any flare of QPE in 
the scenario of star-disc collision. 

We find that the long-term orbital periodic evolution strongly depends on the surface density of the disc ($\Sigma$, which is determined by the accretion disc model) and the orbital inclination of the orbiter $I$. 
The summaries of the simulation results can be found in Tables~\ref{tab:model_gsn069}
and~\ref{tab:model_eroqpe2} for GSN069 and eRO-QPE2, respectively. 
{The initial value of other orbital elements will not affect 
the result of the periodic decay. Thus, without specification here we fixed them to be 
$\Omega=30^\circ$, $\omega=120^\circ$ and $f=180^\circ$. As these parameters determine the orbital 
phase or rotations on the orbital plane, setting them to different values mainly 
result in a different oscillation pattern of the timing of QPE.}
The details of the results and comparisons with the observations are shown in the following sections.

\subsection{The case of an SBH orbiter}
\label{subsec:BH_orbiter}
If the orbiter is a $10\msun$ SBH, the simulations show that
the decay rate of the orbital periods is in the order  of $\dot P\sim -0.1\sim 10$ s\,yr$^{-1}$, as shown in Tables~\ref{tab:model_gsn069} and~\ref{tab:model_eroqpe2} for GSN069 and eRO-QPE2, 
respectively. We find that these results are well consistent with those in Equation~\ref{eq:dotp_grav}. 
As the inclination angle is already assumed to be very high in
these simulations, the values can be considered as the upper limits of the model.

Another important mechanism of orbital decay for SBH
is gravitational wave radiations, which occurs within a timescale of~\citep{1964PhRv..136.1224P}
\be\ba
T_{\rm GW} &\simeq 10^7{\rm\,yr}\times\frac{10^{-5}}{\nu}
\left(\frac{a}{300\, r_g}\right)^4 \frac{\bh}{10^6\,\msun}f(e)\\
f(e)&=\frac{(1-e^2)^{7/2}}{1+\frac{73}{24}e^2+\frac{37}{96}e^4}
\ea\ee
where $\nu=M_\star/\bh$ is the mass ratio between the orbiter and the MBH. 
The expected decay rate of the orbital period  is then given by 
\be
\dot P_{\rm GW}\sim \frac{P}{T_{\rm GW}}\simeq 
0.02\,{\rm s/yr}\times \frac{\nu}{10^{-5}}\left(\frac{a}{300\,r_g}\right)^{-5/2} f^{-1}(e)
\ee
For GSN069 or ero-QPE2, the orbital decay by this mechanism should be
$\lesssim 2\,{\rm s}$ per year given the mass of SBH $<100\msun$.

Both orbital decay due to gravitational wave radiation and gas drags
are at least $2-3$ orders of magnitude lower than those of GSN069 or eRO-QPE2. 
Thus, SBHs are unlikely  the orbiters for QPEs in GSN069 and eRO-QPE2.

\subsection{The case of a WD orbiter}
\label{subsec:WD_orbiter}
The simulation results of the
orbital decay of a WD orbiter are shown in Tables~\ref{tab:model_gsn069} and~\ref{tab:model_eroqpe2} for GSN069 and
eRO-QPE2, respectively. We find that the simulated decay of periods
are also consistent well with those analytical values given by Equation~\ref{eq:dotp_fri}.

For GSN069, $\dot P$ is about $<5\sim17$s yr$^{-1}$. Thus, WD is unlikely the orbiter in GSN069.
For eRO-QPE2, $\dot P$ is $\sim-167$s yr$^{-1}$  in model WD3, in the case of PDM with a disc mass of $4\msun$. 
Thus, in principle, it can be up to $-378$s yr$^{-1}$ for a disc mass of $9\msun$. 
However, it should be a rare event as the mass of the progenitor star that forms the disc should be at least $18\msun$. 
In this case, it is required that the surface density of the disc remains at a level of $\sim 10^8$\,g\,cm$^{-2}$ throughout the observed span of $3$\,yr.

\subsection{The case of an MS orbiter}
\label{subsec:MS_orbiter}
\begin{figure*}
\centering  
    \includegraphics[scale=0.7]{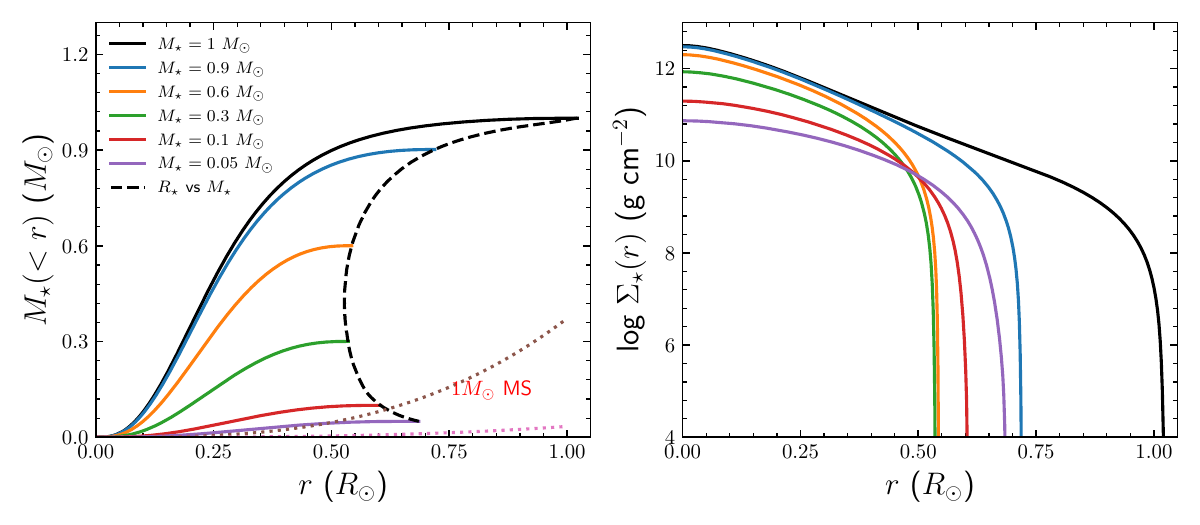}
    \includegraphics[scale=0.7]{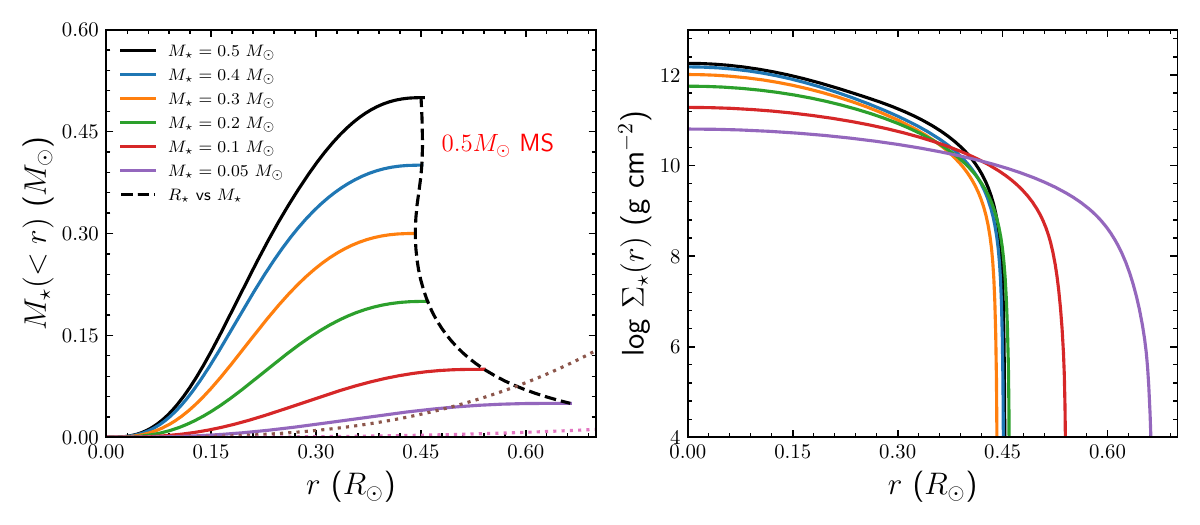}
    \caption{Top panels: The enclosed mass and the surface density profile of a $1\msun$ MS
    star  obtained from MESA at its age of $4.5$Gyr. The black dashed line in the left panels 
    shows the adiabatic response of the radius when the star loses its mass by gradual 
    ablation of the outer envelope. {
    The brown dotted line show the criteria 
    below which the MS is tidally disrupted in eRO-QPE2, i.e., 
    at a distance of $300r_g$ given $10^5\msun$ MBH. The pink dotted line 
    is similar but for MS in GSN069 at a distance of $320r_g$ given  $3\times10^5\msun$ MBH.
    }
    Bottom panels: Similar to the top panels but for an MS of mass $0.5\msun$.}
    \label{fig:den}
\end{figure*}

\begin{figure*}
    \centering
    \includegraphics[scale=0.7]{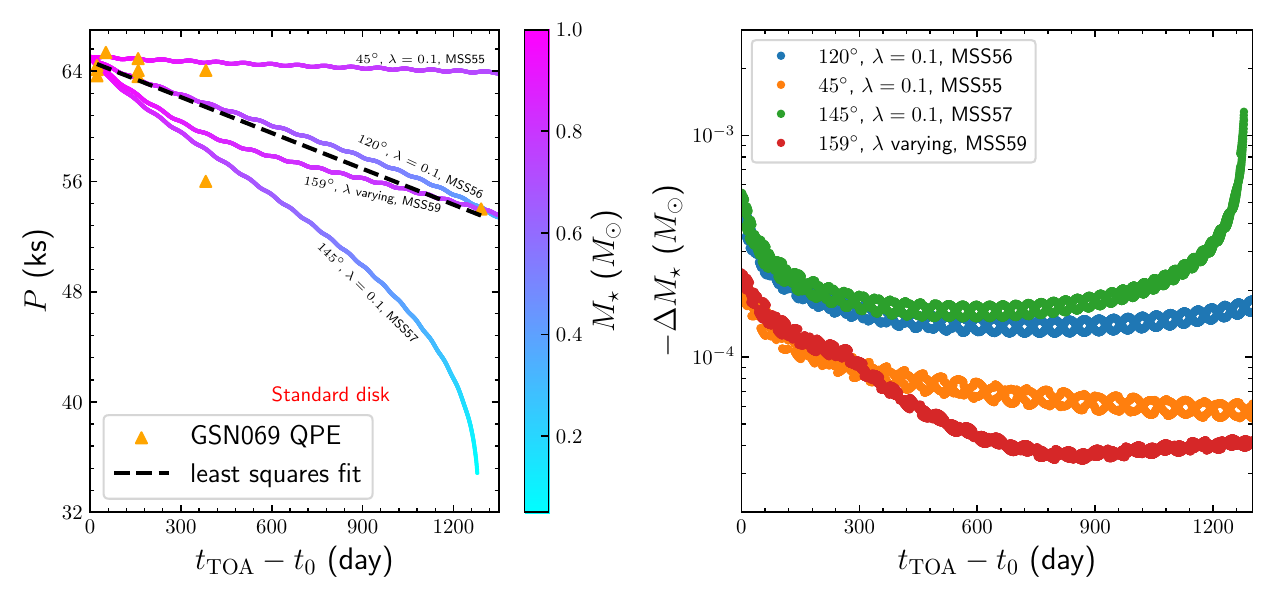}
    \includegraphics[scale=0.7]{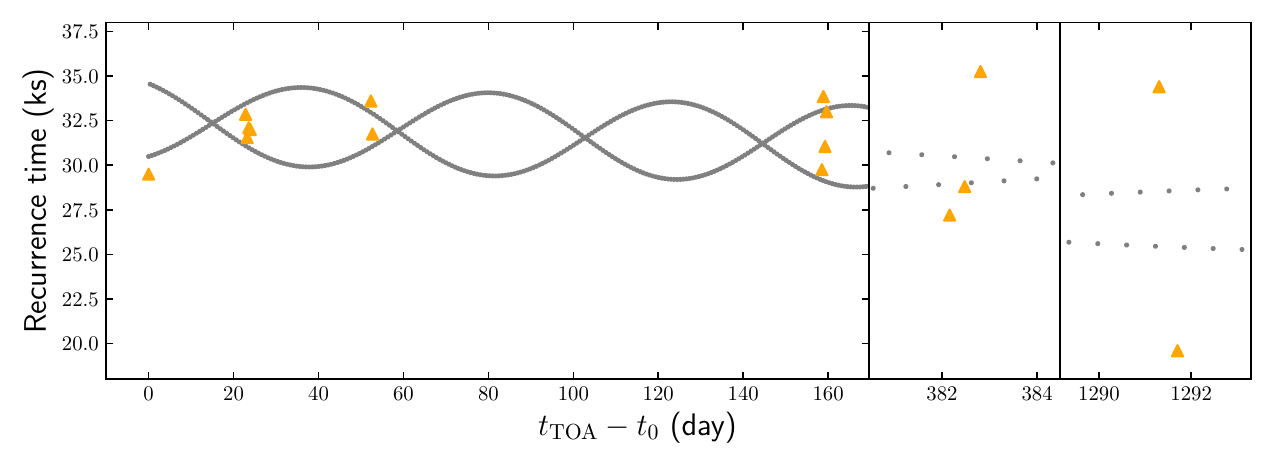}
    \caption{Top left panel: The simulated evolution of the periods of QPEs in GSN069 assuming 
    SDM and a solar-mass MS orbiter. 
    The horizontal axis is the time with respect to the first detection 
    of the QPE in GSN069 (Dec. 24, 2018, see Table~\ref{tab:obs_data_gsn069}).
    The color of each line represents the remaining mass of the MS, where the mapping 
    is shown on the right of the panel. Lines from the top to the bottom 
    show the evolution in models MSS55, MSS56, MSS59 and MSS57 in Table~\ref{tab:model_gsn069}, respectively.
    {The black dashed line represents the least squares fitting result of the observed data in Table~\ref{tab:model_gsn069}
    ($\dot P=-3160\pm 720$\,s\,yr$^{-1}$).}
    Top right panel: The stripped mass after each collision for models on the left panel.
    Bottom panel: Comparison of the simulated (model MS5S9) and the observed time of recurrence in GSN069. 
    }
    \label{fig:gsn069_mass_strip_standard}
\end{figure*}

\begin{figure*}
    \centering
    \includegraphics[scale=0.7]{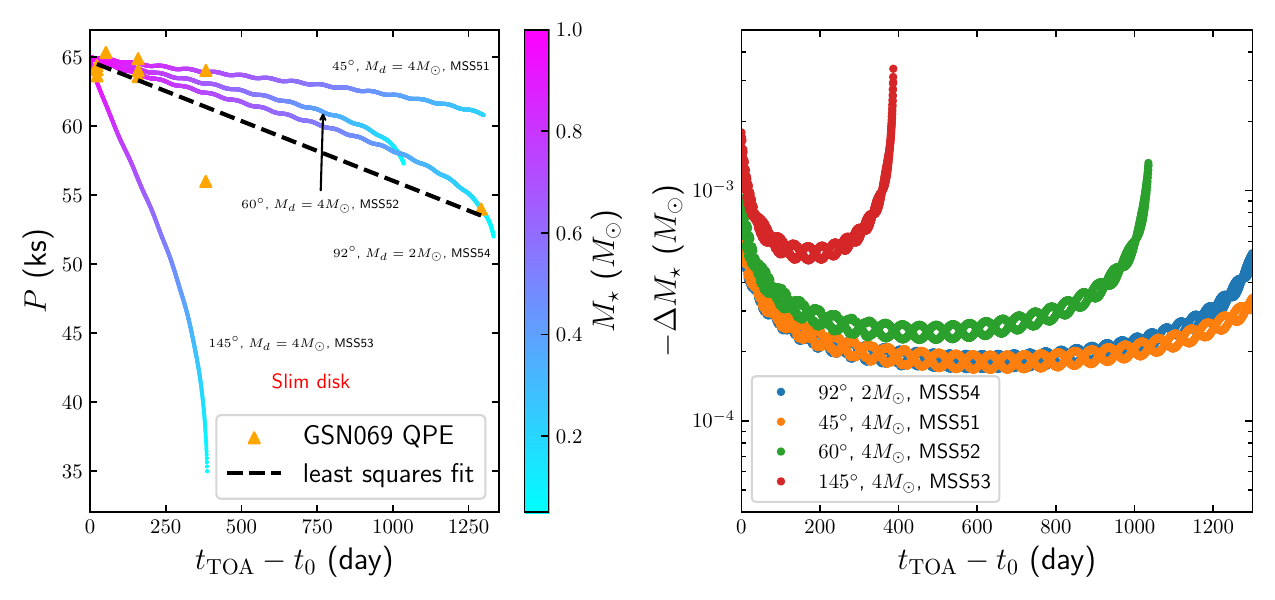}
    \includegraphics[scale=0.7]{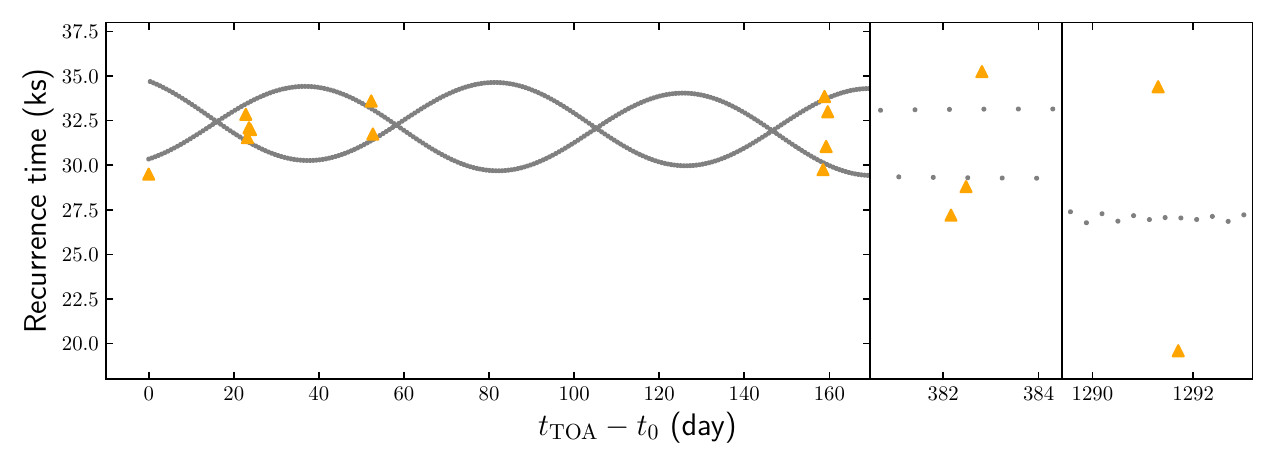}
    \caption{Similar to Figure~\ref{fig:gsn069_mass_strip_standard} but for the case assuming PDM.
    The colored lines in the top left panel, from top to bottom, show the results of models MSS51, MSS52, MSS54, 
    and MSS53, respectively. The bottom panel shows the simulated recurring time of model MSS54.
    }
    \label{fig:gsn069_mass_strip_slimdisk}
\end{figure*}

Another possibility is that the orbiter is a stellar orbiter, such as an MS~\citep[e.g.,][]{1987ApJ...318..794H,2021ApJ...917...43S,2024arXiv240714578Y}. 
Due to the supersonic shocks, the outer envelope of the MS is expected
to be gradually stripped each time it collides with the gas on the disc. To
consider the mass stripping we need an accurate stellar density profile as a function of radius. 
Here we adopt the sophisticated stellar evolution code Modules for Experiments in Stellar 
Astrophysics~\citep[MESA]{Paxton2011, Paxton2013, Paxton2015, Paxton2018, Paxton2019, Jermyn2023} 
(See Appendix for its brief descriptions)  to obtain the density distribution of $0.5\msun$ or $1\msun$
 MS star at its age of $4.5$\,Gyr. As the time interval between successive collisions with the disc is expected to be comparable to the dynamical timescale of a star but orders of magnitude shorter than its thermal timescale, the star reacts adiabatically to the mass loss in its outer
envelope~\citep[e.g.,][]{1987ApJ...318..794H,2024arXiv240714578Y,2024arXiv240601670L}.
We then gradually strip the outer envelope of the star by adopting a wind loss rate of $10^{-6}\sim 10^{-5}\msun$ yr$^{-1}$, following procedures similar to 
those in~\citet {2024arXiv240601670L}. This artificial wind
is used to obtain the adiabatic response of the star's envelope due to mass loss.
The resulting mass and density profiles of the stripped stars are shown
in Figure~\ref{fig:den}.

The striping of the envelope of the star due to its collision with the disc gas is considered following a scenario similar to those in~\citet{1975ApJ...200..145W} 
and~\citet{1996ApJ...470..237A}. 
In the frame of the MS, the impact momentum of the gas from the disc can blow away 
gas outside some critical radius $R_{\rm crit}$ (by accelerating them 
to a velocity larger than its local escape velocity). Using the surface density $\Sigma_\star$
of the star obtained from MESA, we can obtain the critical radius $R_{\rm crit}$ by solving equation~\citep{1975ApJ...200..145W}:
\be
\Sigma_\star (R_{\rm crit})=\frac{v_{\rm rel}}{v_{\rm es}(R_{\rm crit})}\Sigma
\ee
where $v_{\rm es}(R)=\sqrt{2GM_\star(<R)/R}$ is the stellar escape velocity at distance $R$ from the center, 
$\Sigma$ is the surface density of the disc, $v_{\rm rel}$ is the relative velocity between
the star and the gas on the disc. 

The impacting gas can also cause ablation of the gas in regions with a projected distance less than $R_{\rm crit}$~\citep{1975ApJ...200..145W}:
\be
M_{\rm ab}\simeq \Sigma \pi R_{\rm crit}^2 \left(\frac{v_{\rm rel}}{v_{\rm es}(R_{\rm crit})}-1\right) 
\ee
The above analytical estimates of the ablated mass
are already found to be consistent well with numerical
simulations in~\citet{1996ApJ...470..237A}.

Then after each collision, we can update the mass and radius of the MS to 
$M_\star\rightarrow M_\star(<R_{\rm crit})-M_{\rm ab}$ and 
$R_\star\rightarrow R_\star(M_\star)$ where $R_\star(M)$ is the inverse function of
the enclosed mass $M_\star(<r)$. 

From Figure~\ref{fig:den} we can see that for a $1\msun$ MS the stellar radius first decreases
in response to the decreasing of its mass if $M_\star\gtrsim 0.4\msun$. 
Below $M_\star\sim 0.4\msun$ the stellar radius begins to increase even
if its mass keeps decreasing. 
The radius of a $0.5\msun$ MS remains almost the same above $\sim 0.2\msun$. Below,
the mass of the star begins to expand with a decrease in its mass.

{
In GSN069, the period at the last observation is $\sim 54$ks, corresponding to 
a distance of $\sim 320r_g$ given $\bh=3\times10^5\msun$. From Figure~\ref{fig:den} we can 
see that an MS with 
$0.05\msun$ and at distance $320r_g$ is still above the tidal radius.
As the density of the MS is quite low below $0.05M_\odot$, we 
expect that it is quickly dstroyed within the next few collisions with the disc.
Thus, for GSN069, we stop the star-disc collision if $M_\star<0.05\msun$.
}

{
For eRO-QPE2, we find that the period at the last observation ($\sim 16.3$ks) corresponds to 
a distance of $\sim 300r_g$ given $\bh= 10^5\msun$. At such a distance, 
Figure~\ref{fig:den} show that when an initially $1\msun$ MS reduces its mass to $<0.088\msun$ (or an 
initially $0.5\msun$ reduces its mass to $0.075\msun$), it is tidally disrupted before destroyed completely 
by the collision with the disc.
Thus, we stop the simulation of eRO-QPE2 if its mass is below these critial values.
}

We then set up simulations assuming MS orbiters in GSN069 and eRO-QPE2 (mainly focused on the case of 
$1\msun$ MS). The results of some typical models are summarized in Tables~\ref{tab:model_gsn069}
and~\ref{tab:model_eroqpe2}. In the following Sections~\ref{subsec:gsn069_standard}-\ref{subsec:ero_qpe_stellar} 
we mainly focused on the results for a $1\msun$ MS. The difference of adopting an MS
of $0.5\msun$ or other masses will be discussed in Section~\ref{subsec:subsolarMS}. 

\subsubsection{For GSN-069}
\label{subsec:gsn069_standard}
\begin{figure*}
    \centering
    \includegraphics[scale=0.7]{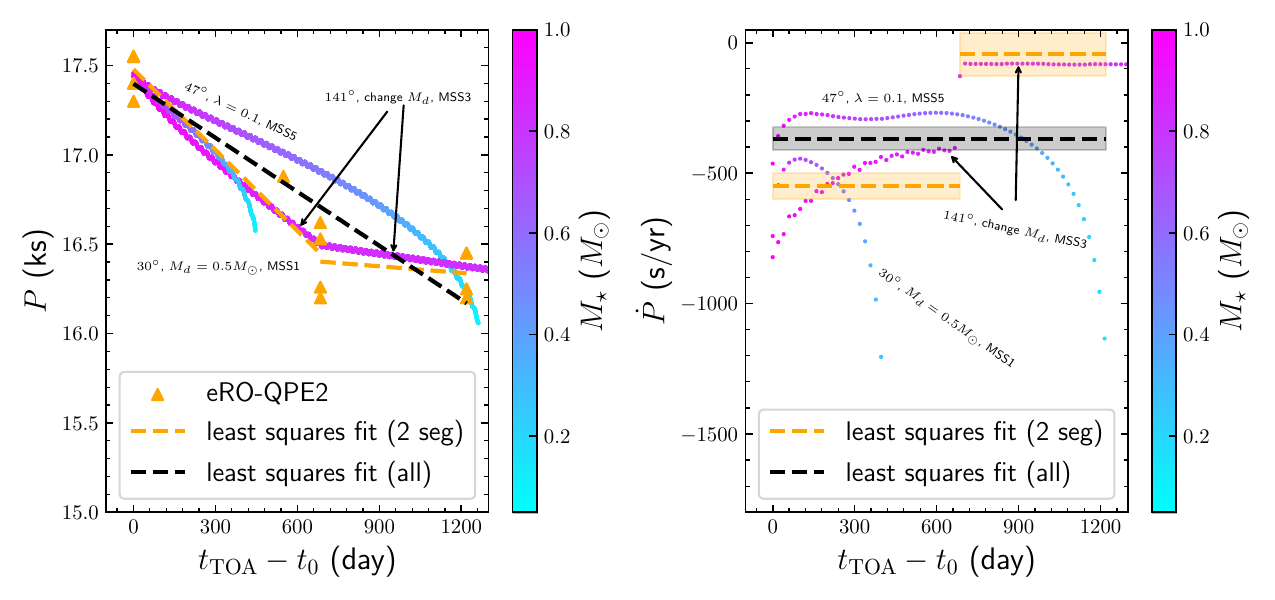}
    \includegraphics[scale=0.7]{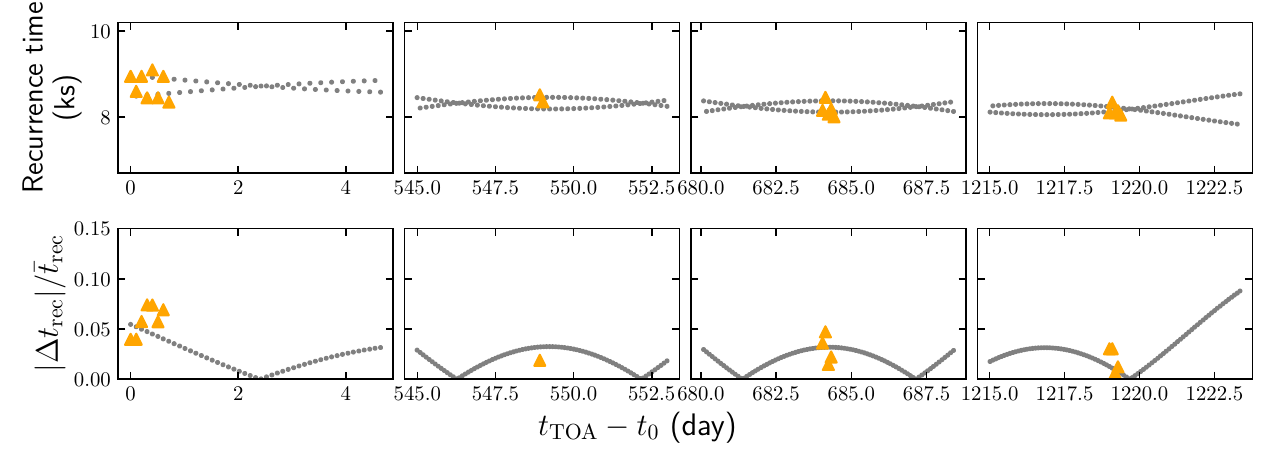}
    \caption{{Top left panel: Similar to the left panel of Figure~\ref{fig:gsn069_mass_strip_standard}, but 
    for source eRO-QPE2. The colored lines show the results of model 
    MSS5, MSS3, and MSS1 in Table~\ref{tab:model_eroqpe2}. Top right panel: The evolution of decay rate of orbital period ($\dot P$).
    $\dot P$ is evaluated from the current moment over a time interval spanning $100$ periods of QPEs. 
    The dotted yellow line and shadow show the estimated mean and error of $\dot P$ by 
    least squares fitting method in the first $685$ days of observations 
    (From Aug., 6th, 2020 to Jun. 21, 2022), and those after that (From Jun. 21, 2022 to Dec. 08, 2023). 
    In both of the top panels the black dashed lines show the least squares fitting result of
    all data in Table~\ref{tab:obs_data_eroqpe2} ($\dot P=-370\pm40$\,s\,yr$^{-1}$). 
    Bottom panel: The recurrence time and the contrast of recurring time 
    $|\Delta t_{\rm rec}|/\bar t_{\rm rec}$ in model MSS3, where $|\Delta t_{\rm rec}|$ and 
    $\bar t_{\rm rec}$ is the difference and mean value 
    of the recurring time between successive QPEs. Initially we 
    set $\omega=345^\circ$ and $f=200^\circ$ such that the model has good consistencies 
    with the observation.}}
    \label{fig:ero_QPE2_mass_strip}
\end{figure*}
As shown in Figure~\ref{fig:gsn069_mass_strip_standard},
in the case of SDM, it is possible to
reproduce the required decay of $\sim -3160\pm720$ s\,yr$^{-1}$ 
if fixing $\lambda=0.1$ and given a large inclination angle of collision, i.e. $I= 120^\circ$ (model MSS56). 
In this case, the surface density of the disc is around $10^6$\,g\,cm$^{-2}$. 
The mass ablation of the star per collision begins with 
$\Delta M_\star\sim-4\times10^{-4}\msun$ 
but quickly settles down to $\sim -2\times10^{-4}\msun$.
The final mass of the star after $\sim 1200$ days is about $0.4\msun$.
The star is expected to be completely destructed at the time of $\sim1660$ day since its first QPE ($\sim 1$yr after 
the last observation of QPE of GSN069 in 2022).

If we vary $\lambda$ according to~\citet{2023A&A...670A..93M}, 
to reproduce the observed orbital decay, we find that the inclination
angle needs to be slightly larger, i.e., $I\sim 159^\circ$ (model MSS59), suggesting that the orbiter is cycling the MBH in a nearly retrograde orbit. 
As the inferred $\lambda$ in GSN069 varies between $0.6\sim 1.3$~\citep{2023A&A...670A..93M}, 
the surface density of the disc felt by the orbiter in GSN069 becomes relatively lower ($3\sim 6\times10^5$ g\,cm$^{-2}$) in SDM (see Figure~\ref{fig:par_disk}).
The ablation of mass in this model is on levels of $-2\times10^{-4}\msun\sim-3\times10^{-5}\msun$ per collision. 
The final mass of the star after $\sim 1200$ days is about $0.76\msun$ and the
star can continue to produce QPEs until $\sim 3600$ days ($\sim 10$ yr) since the first QPE.

The above two models assume an MBH mass of $3\times10^5\msun$. However, the results remain similar if alternatively we assume an MBH mass of 
$10^6\msun$, e.g., in model MSS62 and MSS63. This is mainly because both the orbital decay (see Equation~\ref{eq:dotp_fri}) and the mass stripping 
is not or weakly depends on the mass of MBH.

In both of the above two models (MSS56 and MSS59), the recurring patterns of QPEs 
can be consistent with the observations of XMM3, XMM4, Chandra (22096) and XMM5, 
which covers a time span of about $160$ days with respect to the first flare (see the bottom panel
of Figure~\ref{fig:gsn069_mass_strip_standard} for the case of model MSS59).
However, after that, the recurring patterns show some discrepancies between the model predictions
and observations, which can be about $\sim 5$\,ks when the recurrence times are about $25\sim 30$\,ks. 
In other cases of the orbital configurations, e.g., MSS55, MSS57-8, 
the resulting orbital decays are inconsistent with the observation, which is either too large or small.

The irregular timing patterns possibly because, within a short duration of about $200$ days, 
the disc is significantly disturbed (e.g., becomes warped) due to the injection of gas streams from another tidal disruption event~\citep{2023A&A...670A..93M,2024PhRvD.109j3031Z}, 
and/or that the orbiter has increased its eccentricity to $\sim 0.2$  
by some unknown mechanisms. However, these complexities should mainly affect
the timing pattern of QPEs. Our resulting decay of orbital period should remain somewhat accurate, 
as long as the surface density and impact angle felt by the orbiter per orbit does not change
significantly due to these complexities, 

In the case of PDM, the density of the disk in GSN069 is mainly regulated by the assumed total mass of the disk 
(or the assumed mass of the progenitor star that is disrupted). 
If assuming a $4\msun$ disk mass, the surface density felt by the star is
$4\times10^6$\,g\,cm$^{-2}$, about four times higher than in the case of SDM adopting $\lambda=0.1$. 
As shown in Table~\ref{tab:obs_data_gsn069} and Figure~\ref{fig:gsn069_mass_strip_slimdisk}, 
it is possible to reproduce a decay rate of $\dot P\sim 2700$\,s\,yr$^{-1}$. 
However, in a such high density the mass stripping of the star is so significant that 
it is quickly destroyed within $\sim1000$ days (model MSS52, see also top left panel of Figure~\ref{fig:gsn069_mass_strip_slimdisk}). 

Thus, to reproduce the observations, the disk mass should be smaller, e.g., $\le 2\msun$.
An example of such cases is shown in model MSS54 which assumes $M_d=2\msun$, although the 
star is quickly destroyed within $1330$ days. Thus, in order to survive for $\gtrsim 3.5$\,yr, 
the upper limit of surface density of the disc should be $\sim 2\times10^6$\,g\,cm$^{-2}$. If the disk mass is less than $2\msun$, 
the MS can survive for a longer time, which is similar to those cases of SDM mentioned in the previous Section.

According to Table~\ref{tab:model_gsn069}, we can see that in low surface density ($3\sim6\times10^5$\,g\,cm$^{-2}$), to 
reproduce the observed decay of period the inclination angle is already high ($146\sim 160^\circ$ in model MSS61, MSS63 and MSS59). 
Discs with surface density lower than these values require inclination angles $>160^\circ$ in order to reproduce the observed decay of periods. 
However, orbiters in such nearly retrograde orbits are not expected to produce QPE flares, as they are completely trapped within the 
gas in the disc. Thus, these results suggest that the lower limit of surface density of the disc is $\sim 3\times10^5$\,g\,cm$^{-2}$, for a $1\msun$ 
MS to reproduce the observed decay of GSN069.

\subsubsection{For eRO-QPE2}
\label{subsec:ero_qpe_stellar}
The evolution of the orbital period in the case of an MS orbiter in eRO-QPE2 can be found
in Table~\ref{tab:model_eroqpe2} and Figure~\ref{fig:ero_QPE2_mass_strip}. In the case of SDM, we tested three cases of the 
Eddington ratio: $\lambda=0.01$, $0.1$ and $1$. We find that all of them can
produce $\dot P=-360\sim -390$ s yr$^{-1}$ by tuning the inclination angle $I$ in the range of
$47^\circ\sim 102^\circ$. In the case of $\lambda=0.1$ ($\Sigma \sim 10^6$ g cm$^{-2}$ and $I=47^\circ$), 
the star is completely disrupted within $\sim 1300$ days, suggesting that QPEs are no longer observable
after the last observations of QPEs in eRO-QPE2 (Dec., 2023). In other two cases ($\lambda=0.01$ or $\lambda=1$) the 
surface density of the disc is lower ($\Sigma\sim 3\times10^5$) and thus they can have longer survival time.
These results suggest that, to reproduce the observed decay rate of eRO-QPE2
the disc surface density is required to be $\lesssim 10^6$ g cm$^{-2}$.

In PDM, we can explore the surface density of the disc beyond the SDM. We find that
if $M_d=0.5\msun$ (MSS1), the disc surface density is so high ($\Sigma\sim 5\times10^6$ g cm$^{-2}$) 
that the MS is completely destroyed after $\sim 450$ days even if the inclination is low $\sim 30^\circ$. 
We also test the case $M_d=0.01\msun$ (MSS2), a mass suggested by~\citet{2023A&A...675A.100F}. 
In this case, it has a decay rate of $\sim 380$\,s\,yr$^{-1}$, which consists with those of the observed value of $370\pm 40$\,s\,yr$^{-1}$.
The density is relatively low ($\Sigma\sim 10^5$ g cm$^{-2}$), such that the MS can
last for more than $12$yr ($4569$ days). Considering that the inclination angle is already high 
($135^\circ$), to reproduce the observed periodic decay the surface density of 
the disc can not be lower than $\sim 5\times10^4$ g cm$^{-2}$.

{Although the above simulations can reproduce the observed decay of orbital periods in a time span of $\sim 3.5$ yr, 
from Figure~\ref{fig:ero_QPE2_mass_strip} it seems that they lower-predict 
the decay of the period before the third observation (Jun. 21, 2022) of eRO-QPE2 ($\sim 684$ days after the 
first observation of eRO-QPE2), and over-predict after that. As shown in the top right panel of Figure~\ref{fig:ero_QPE2_mass_strip}, 
the decay of the orbital period is around $-550\pm51$ s\,yr$^{-1}$ in the first three observations and 
reduce to $-45\pm 83$ s\,yr$^{-1}$ between the last two observations. This result suggests, $684$ days after the first observation, 
the decay of the orbital period is $\gtrsim 4$ times smaller.}

{Here we discuss three scenarios that can possibly cause the slowing down of the orbital decay. 
The first one is that the surface density of the disc felt by the orbiter
decreases as it moves inwards. Note that in this scenario we assume that 
the surface density of the disc remain unchanged during the time of observation. As the orbital periods 
reduce from $\sim 17.5$ks to $\sim 16.3$ks in the observed span of $\sim 3.5$yr, the distance 
of the orbiter to the MBH reduce from $\sim 316r_g$ to $\sim 300r_g$, assuming an MBH of mass $10^5\msun$ for eRO-QEP2.
According to Equation~\ref{eq:dotp_fri}, we have $\dot P\propto \Sigma$. Thus, to decrease 
the surface density of the disc by a factor of $\gtrsim 4$ when $r$ moves from $316r_g$ to $300r_g$,  
it is required that $\Sigma\propto r^{\gamma}$ and $\gamma\gtrsim 27$. 
Such a dramatic increase of the density along the radius seems very unlikely for a disc model.}

{According to Equation~\ref{eq:dotp_fri} we have $\dot P\propto R_\star^2m_\star^{-1}$, and thus 
the second possibilities  may be the combination effects of the changes 
of the mass and radius of the MS. However, from Figure~\ref{fig:den} we can see that for an $1\msun$ MS 
$\dot P$ can only decrease by a factor of $\lesssim 2$, of which the maximum occurs 
when $M_\star\sim 0.5\msun$ and $R_\star\sim 0.5R_\odot$. The top right panel of Figure~\ref{fig:ero_QPE2_mass_strip}
also show the simulated $\dot P$ for model MSS5. We can see that in that model $\dot P$ decreases with the mass 
of the MS if $M_\star\gtrsim 0.5\msun$, however, by only a factor of about $\sim 1.7$, which is still far below the
minimum required factor of $4$. The decline of $\dot P$ in the model occurs rapidly at the beginning and 
finished within the $\sim 50$\, days. However, 
observations require that it should happen around $680$\,day after the first observation. 
}

{Finally, it is possible that the surface density of the disc felt by the orbiter is decreasing with time. For example, 
the decrease of the total mass of the disk, the shrinking of the outer 
edge of the disc to the regions near the orbiter such that it feels much lower density, etc.  
Here we do not intend to delve deeply to the physical mechanisms behind such a decreasing of surface density with time, 
but explore whether it can result in $\dot P$ and the recurring time patterns consistent with those of the observations.
A simple way to realize such a scenario is shown in model MSS3 in Table~\ref{tab:model_eroqpe2}, i.e., 
a PDM begins with a disc mass $M_d=0.01\msun$  but after $684$\,days it reduces to $5$ times lower 
($M_d=2\times10^{-3}\msun$).
The resulting period (and the decay $\dot P$) is generally consistent with the observation, as 
shown in the top panels of Figure~\ref{fig:ero_QPE2_mass_strip}.} 

{We also compare the recurring time of QPE in model MSS3 with those in observations in the bottom panels of 
Figure~\ref{fig:ero_QPE2_mass_strip}. We define 
$|\Delta t_{\rm rec}|/\bar t_{\rm rec}$ 
as the contrast of the recurring time, where $|\Delta t_{\rm rec}|$ and $\bar t_{\rm rec}$ 
is the difference and the mean of the recurring time in successive QPE flares, respectively. 
We can see that both the recurring time and contrast can be consistent with the observations.
}

\subsubsection{The case of MS with other masses}
\label{subsec:subsolarMS}
The results in Sections~\ref{subsec:gsn069_standard}-\ref{subsec:ero_qpe_stellar} apply only for a solar-mass MS. 
If the star is less massive, i.e., $0.5\msun$, we find that it can also reproduce the observed decay of GSN069 or eRO-QPE2, 
as shown in Tables~\ref{tab:model_gsn069} and~\ref{tab:model_eroqpe2} (model MSL series), 

The mass stripping of a $0.5\msun$ MS is also in the order of $10^{-4}\sim10^{-5}\msun$ per collision.
Given the same condition of the disc, the rate of mass stripping is slightly higher than the case of a $1\msun$ MS, 
as its mean density is relatively lower. Mainly because of its smaller mass, the survival time of the star is in general shorter than in
the case of a solar-mass MS. According to Equation~\ref{eq:dotp_fri}, it seems that it requires a larger impact angle in order to produce 
the same amount of orbital decay. However, as its mass decreases, $\dot P$ can be accelerated to larger values due to its fast expansion of the radius. Thus, in general, the impact angle needed to reproduce the observed decay of GSN069 or eRO-QPE2 in a $0.5\msun$ MS is more or less similar to the case of a $1\msun$ MS.

If the initial mass is smaller, i.e., $<0.5\msun$, the MS is expected to be more vulnerable to collisions, and thus the survival time is even shorter. For GSN069 we find that such a small mass MS is unlikely, 
as either its orbit is required to be extremely retrograde (close to $180^\circ$), or that the required surface density of the disc is so high that its survival time becomes shorter than $\sim3$yr. For MS with initial masses $>1\msun$, by varying some conditions of the disc and the impact angle, it is possible to reproduce the observed decay of periods both for GSN069 or eRO-QPE2, similar to the case of a $1\msun$ MS. However, since the expected numbers of them are smaller and their lifetime is shorter than those of a $1\msun$ MS, the possibility of placing them around MBH in orbits of $10^2r_g-10^3r_g$ is less likely.

\subsection{The case of a naked Helium core}
\label{subsec:Hecore_orbiter}
\begin{figure}
\centering
    \includegraphics[scale=0.7]{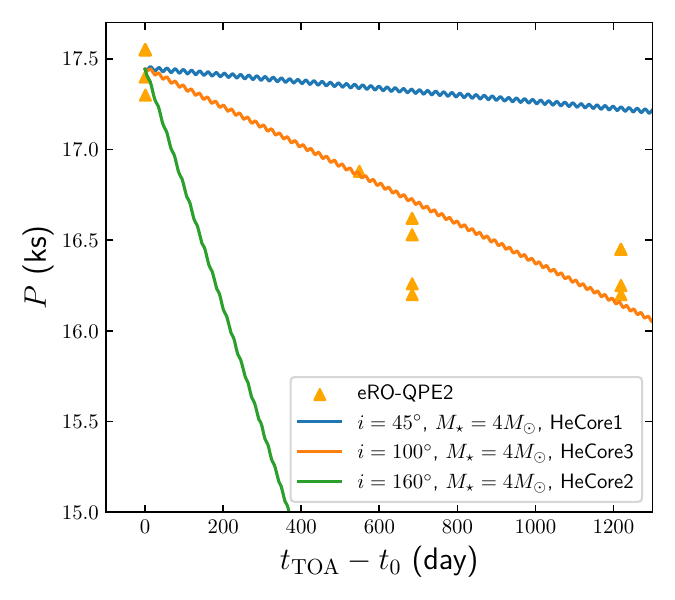}
    \caption{{The evolution of the simulated periods of QPEs in eRO-QPE2 assuming a naked Helium 
    core orbiter and PDM disc models. The horizontal axis is the time with respect to the first detection of the QPE in eRO-QPE2 
    (August, 6th, 2020 in Table 2). Lines with different colors represents the results of model 
    HeCore1-3 in Table~\ref{tab:model_eroqpe2}.}}
    \label{fig:ero-QPE2_core}
\end{figure}
In this section, we discuss the orbital evolution of the orbiter if it is
a naked Helium core. Such an object may form near the MBH through a previous
partial TDE of a red giant~\citep[e.g.,][]{2021ApJ...921L..32X}. Its outer envelope is removed
either through repeating partial TDE or repeating collisions with the accretion disc form by the partial TDE.
Such an object can be migrated into the orbits suggested by GSN069 or eRO-QPE2 
through two-body dynamic relaxations. Note that the partial TDE of the red giant can be a distinct event 
occurred long before the TDE forming the  disc that the remnant core is current colliding with.

Following~\citet{2012ApJ...757..134M}, here we adopt a red giant of mass $1.4\msun$ and
use MESA to evolve such a star until its age is $3.56$Gyr.
Then we adopt a similar process described in Section~\ref{subsec:MS_orbiter}
to remove the outer envelope until a naked Helium core is left. We find that
such a core is of size $\sim 0.03R_\odot$ and mass $\sim 0.26\msun$.

We perform simulations assuming that such a Helium core is the orbiter in GSN069 and eRO-QPE2. 
The results of the evolution of the period can be found in Tables~\ref{tab:model_gsn069} and~\ref{tab:model_eroqpe2}.
We can see that the Helium core is not favored in GSN069 as the decay 
rate of the orbital period of the remnant core is in the order  of $\dot P\lesssim 200$ s yr$^{-1}$, 
about one order of magnitude lower than those expected from observations.

For eRO-QPE2, we find that it is possible to reproduce the observed decay rate; however, the surface density of the disc is required to be larger than $10^7$ g\, cm$^{-2}$ 
(see model HeCore3 or HeCore4). 
This can be realized assuming the disk mass $\gtrsim1\msun$ 
in PDM. The evolution of the orbital period in model HeCore1-3 can be found in Figure~\ref{fig:ero-QPE2_core}. 
However, in the SDM model, since the surface density is $\lesssim 10^6$\,g\,cm$^{-2}$, it is
difficult to reproduce the period decay rate in eRO-QPE2.

\section{Discussion}
\label{sec:discussion}
In this study, we have explored the long-term periodic evolution of the star-disc system under two assumed accretion disc models. However, considering that there are large uncertainties in the accretion physics, we do not intend to rule out any accretion disc model. 
In fact, as suggested in Equation~\ref{eq:dotp_fri}, for a given type of orbiter the long-term orbital decay is mainly determined by the surface density of the disc and the impact angle.
The density profile of the disc is only important if the orbital decay is so significant that 
it moves to the inner parts of the disc. We find that within the observed span of $\sim 3$yr, 
surface density of the disc felt by the orbiter in both GSN069 and eRO-QPE2 is not changed significantly. For example, 
The orbital semi-major axis of the orbiter in GSN069 (eRO-QPE2) is changed from $365r_g$ to $\sim 320r_g$ 
($316r_g$ to $\sim 300r_g$). Thus, our simulation results are more sensitive to the 
value of the surface density of the disc rather than the disc models assumed. 

Nevertheless, the surface density suggested by QPE orbiters obtained from our study 
may help to speculate the underlying accretion disc model. For example, 
if eRO-QPE2 has a Helium core orbiter, PDM is more likely as it is possible to have a much higher surface 
density than those in SDM, as discussed in Section~\ref{subsec:Hecore_orbiter}.


We have already shown in Section~\ref{subsec:MS_orbiter} that a solar-like MS orbiter explored here can reproduce the orbital decay of both GSN069 and eRO-QPE2. Note that 
the mass stripping is assumed to begin at the time of the first QPE observed in these sources. 
However, this may not be true for both GSN069 or eRO-QPE2. 
For example, it seems that the onset of the active
phase of GSN069, which indicates the existence of the disc in our simulation, 
should be $10\sim26$ years before its first detection of QPE~\citep{2011arXiv1106.3507S, 2023A&A...670A..93M} 
in 2018 (see Table~\ref{tab:obs_data_gsn069}). Before the first eruption, eRO-QPE2 was previously identified as a star-forming galaxy and was not detected in X-rays up to 2013 by ROSAT and XMM-Newton slew
survey~\citep{2021Natur.592..704A}. However, as the time of exposure of these two surveys are 
much shorter than the recurring time of QPE in eRO-QPE2, it is likely that they have 
missed the flare even if it was active during that period. 
Thus, if the first collision of the MS with the disc is before
the first detection of QPEs, the mass of the star is expected to be larger than $0.5\sim1\msun$.

For GSN069, the last measured period in Table~\ref{tab:obs_data_gsn069} is speculated
according to the correlation between QPE and the oscillation pattern of the quiescence flux~\citep{2023A&A...674L...1M}. 
Thus, conservatively, there should be only a lower limit of the period at that time. 
If the true period is indeed much larger than this value, there is no obvious trend of secular evolution of QPEs. Thus, in principle, all orbiters are possible in GSN069. However, for MS, we
still require it to exist for more than $\sim 3.5$ yr. Thus, the requirement of the
upper limit of the surface density of the disc felt by the orbiter, i.e., lower than $2\times10^6$\,g\,cm$^{-2}$ (Section~\ref{subsec:MS_orbiter}), should remain valid.

{For GSN069, from Figure~\ref{fig:gsn069_mass_strip_standard},~\ref{fig:gsn069_mass_strip_slimdisk} it seems that the matching between the simulated and observed period is not satisfactory. 
For example,  the period measured in XMM6 (Jan. 2020) has two distinct periods of $55$ks and $64$ks. 
Such a complexity can not be reproduced by our simple models, suggesting that there are more complexities not included in our study. 
We speculate that future observations of QPEs of GSN069 may be needed to see if these discrepancies are due to complexities
in accretion physics or to fluctuations of the timing of QPE flares.}

\section{Conclusions}
\label{sec:conclusion}
X-ray quasi-periodic eruptions (QPE) are repeating intensive and rapid bursts of X-ray emissions recently observed in the center of some nearby active galaxies. Assuming that QPEs are driven by an orbiter colliding with
the accretion disc around the massive black hole (MBH), i.e., the so-called ``star-disc model'', 
the orbital energy of the orbiter may be gradually dissipated because of its collision with the gas on the disc.
In this study, we set up numerical simulations to investigate the long-term periodic evolutions of the QPEs and compared them with the QPE sources GSN069 and eRO-QPE2, both of which
are observed with evidences of decaying orbital periods. 

The periodic decays of GSN069 and eRO-QPE2 are about $\sim -3160\pm720$\,s\,per year and $\sim -370\pm40$\,s\,per year, 
respectively. We find that for both of these two sources, the orbiter is unlikely to be
a stellar-mass black hole (expected to have a mass of $<100\msun$), as its periodic decay is $<10$\,s\,yr${^{-1}}$,
at least two orders of magnitude lower than the observed values. 

A white dwarf (WD) orbiter is unlikely in source GSN069 as the expected orbital decay is 
$\lesssim 20$s per year. Although WD is possible to reproduce the observed orbital decay in eRO-QPE2,  
it requires a very high surface density of the disc, i.e., $\sim 10^8$ g\, cm$^{-2}$ during the 
observed span of $\sim 3.5$yr.

We find that if the orbiter is a solar-like main-sequence star (MS), the orbital periodic decay
can be consistent with those observed in both GSN069 and eRO-QPE2. However, the MS
will eventually be destroyed because of the continuous ablation of the outer envelope each time
it collides with the disc. If the mass of MS is $1\msun$, for GSN069 the disc surface density 
needs to be $\lesssim2\times10^6$g\,cm$^{-2}$, 
otherwise, the MS will be completely destroyed within the observed span of QPE of $3.5$yr.
Meanwhile, the surface density should be large enough, i.e., $\gtrsim3\times10^5$g\,cm$^{-2}$ 
in order to reproduce the orbital decay observed in periods. In this case, the maximum
survival time of MS is around $10$yr.
Similarly, for eRO-QPE2, we find that the surface density is required to be in
the range of about $5\times10^4$g\,cm$^{-2}$ to $10^6$g\,cm$^{-2}$. The survival time of the star
is expected to be $\lesssim12$yr and is shorter for higher surface density of the disc.
{We find that a disc model reduces its surface density 
by a factor of $\gtrsim4$ lower since Jun. 21, 2022 can provide better explanations 
for both the observed decay of period and the recurring timing pattern of eRO-QEP2.}

If the MS mass is smaller, for example, $0.5\msun$, we find that the 
observed decay of period in both GSN069 and eRO-QPE2 can also be reproduced with a similar 
disc condition and impact angle. However, the star is more vulnerable to collisions with the disc,
and thus its time of survival is shorter than that of a $1\msun$ MS.

We investigate the possibility that the orbiter is a naked Helium core, e.g., resulting from removing
the outer envelope of a red-giant which is partially disrupted by the MBH. We find that it is unlikely the orbiter in GSN069, 
as its orbital decay is on the order of $\lesssim 200$\,s\,yr$^{-1}$. However, it is possible in eRO-QEP2, 
assuming that the surface density of the disc can be as high as $10^7$\,g\,cm$^{-2}$. 

Our results should be important for revealing the underlying driven physical mechanisms of 
the QPEs and the nature of the orbiters in star-disc collision models. For an MS orbiter,
we expect that it may disappear in the next few years. 
Future continuous observations of these QPE sources can help confirm or rule out these possibilities.

\section{Acknowledgments}
\noindent
This work was supported in part by the National Natural Science Foundation of China under grant No. 12273006. This work was also supported in part by the Key Project of the National Natural Science Foundation of China under grant No. 12133004 and 11833007. The simulations in this work are performed partly in the TianHe II National
Supercomputer Center in Guangzhou. 

{Softwares: SAS, CIAO.}

\appendix

\section{Data processing of GSN069 and eRO-QPE2}
\label{sec:data_process}

{Here we briefly describe details of the data processing of QPEs of GSN069 and eRO-QPE2 
shown in Table~\ref{tab:obs_data_gsn069} and Table~\ref{tab:obs_data_eroqpe2}, respectively. 
Each table includes the observation ID (ObsID), the start date of each observation, the delay to the first flare, and the 
interval between adjacent flares. Details of data reduction are given below.}

{The {\bf XMM-Newton } data were reduced with SAS (v. 22.1.0) adopting the latest calibration files updated in Oct. 2024.
Only the EPIC-pn observations were used because of their better S/N ratio than EPIC-MOS observations, except the observation of 0893810501 for eRO-QPE2 which both  EPIC-pn and  EPIC-MOS were used.
We use the task epproc to create the pn events files, emproc to create the MOS events file.
After removing the `bad' pixels, we created the high flaring particle background time intervals  
 with `PATTERN$==$0' in the 10--12 keV band for  EPIC-pn, and `PATTERN$<=$12' in the 10--12 keV band for  EPIC-MOS. 
It resulted in the good exposure times  of  63.3 ks, 141.4 ks, 135.4 ks, 141.0 ks, and 59.2 ks,  for XMM3, XMM4, XMM5, XMM6 and XMM12 observation of GSN069 listed in Table~\ref{tab:obs_data_gsn069}, respectively.
XMM1-4 of eRO-QPE2 in Table~\ref{tab:obs_data_eroqpe2} were 95 ks, 60 ks, 25 ks, and 46 ks, respectively. 
For GSN069, we extracted the source light curves from the source region centered on its optical position with radius $\sim36''$, 
while a background  from a nearby source-free area on the same CCD chip with radius $\sim80''$. Similarly, for eRO-QPE2 the source 
and background radius are $\sim23''$ and $\sim60''$, respectively.  We used a time bin size of 50 seconds  for the 0.2-1 keV 
lightcurves of GSN 069, and  a time bin size 
of 25-50 seconds for the  0.2-2 keV lightcurves of eRO-QPE2. Barycentric correction was applied using the SAS task barycen. }

{The data of  Chandra/ACIS-S  observation for GSN069 were reduced using the CIAO-4.17 software package with the calibration 
database CALDB-4.12.0.  We reprocessed the data using Chandra\_repro task and applied barycentric correction  with the axbary tool. We 
extracted the source from a circle region with radius of $\sim5''$, and the background from a  nearby  source-free circle region with a 
larger radius of $\sim15''$, from the level-2 events file with an exposure time of 72.6 ks.  The light curve in 0.3-1 keV was extracted 
and binned with a time binsize of 50 seconds using the dmextract tool.
}

\section{Implementation of MESA}
The MESA EOS is a blend of OPAL \citep{Rogers2002}, SCVH
\citep{Saumon1995}, FreeEOS \citep{Irwin2004}, HELM \citep{Timmes2000},
PC \citep{Potekhin2010}, and Skye \citep{Jermyn2021} EOSes.

Radiative opacities are primarily from OPAL \citep{Iglesias1993,
Iglesias1996}, with low-temperature data from \citet{Ferguson2005}
and the high-temperature, Compton-scattering dominated regime by
\citet{Poutanen2017}.  Electron conduction opacities are from
\citet{Cassisi2007} and \citet{Blouin2020}.

Nuclear reaction rates are from JINA REACLIB \citep{Cyburt2010}, NACRE \citep{Angulo1999} and
additional tabulated weak reaction rates \citet{Fuller1985, Oda1994,
Langanke2000}.  Screening is included via the prescription of \citet{Chugunov2007}.
Thermal neutrino loss rates are from \citet{Itoh1996}.

\bibliography{sample63}{}
\bibliographystyle{aasjournal}

\end{document}